\documentclass[prb,twocolumn,showpacs,preprintnumbers,amsmath,amssymb]{revtex4}

\usepackage{graphicx}
\usepackage{dcolumn}
\usepackage{bm}

\usepackage{epsfig,float,afterpage,amssymb,wrapfig,psfrag}
\newcommand{\be}{\begin{equation}}
\newcommand{\cc}{\mbox{\scriptsize{c}}}

\newcommand{\ee}{\end{equation}}
\newcommand{\bea}{\begin{eqnarray}}

\newcommand{\eea}{\end{eqnarray}}
\newcommand{\bra}{\langle}

\newcommand{\ket}{\rangle}
\newcommand{\ssz}{\scriptsize}

\newcommand{\w}{\omega}
\newcommand{\K}{\mbox{\scriptsize{K}}}
\newcommand{\m}{\mbox{\scriptsize{m}}}
\newcommand{\wm}{\w_{\m}}

\newcommand{\up}{\uparrow}

\newcommand{\down}{\downarrow}

\DeclareMathAlphabet{\bi}{OML}{cmm}{b}{it}
\newcommand{\kk}{\bi{k}}

\renewcommand{\theequation}{\arabic{section}.\arabic{equation}}
\newcommand{\ra}{\rightarrow}

\newcounter{saveeqn}
\newcommand{\alpheqn}{\setcounter{saveeqn}{\value{equation}}%
\setcounter{equation}{0}%
\addtocounter{saveeqn}{1}%
\renewcommand{\theequation}{\mbox{\arabic{section}.\arabic{saveeqn}\alph{equation}}}%
}
\newcommand{\reseteqn}{\setcounter{equation}{\value{saveeqn}}%
\renewcommand{\theequation}{\arabic{section}.\arabic{equation}}}

\newcommand{\cl}{\chi_{\mbox{\ssz{loc}}}}
\newcommand{\dcl}{\chi''_{\mbox{\ssz{loc}}}}

\newcommand{\qq}{\bi{q}}
\newcommand{\bphi}{\mbox{\boldmath$\phi$}}
\newcommand{\cb}{{\cal B}}

\newcommand{\tkn}{T_K}

\newcommand{\cH}{{H}}
\newcommand{\wot}{\tilde{\w}_0}
\newcommand{\ve}{x}
\newcommand{\vw}{y}
\newcommand{\fb}{\zeta^*}
\newcommand{\ff}{\eta^*}
\renewcommand{\sb}{\w^*}
\newcommand{\ffv}{\eta^*}
\newcommand{\seceq}{\setcounter{equation}{0}}
\renewcommand{\sec}{Sec.\ }

\begin{document}

\title{Kondo physics and dissipation: A numerical renormalization-group approach to Bose-Fermi Kondo models}

\author{Matthew T. Glossop}
 \email{glossop@phys.ufl.edu}
\author{Kevin Ingersent}%

\affiliation{Department of Physics, University of Florida,
Gainesville, Florida 32611-8440, USA}

\date{\today}

\begin{abstract}
We extend the numerical renormalization-group method to treat Bose-Fermi Kondo models (BFKMs) describing a local moment coupled both to a conduction band and to
a dissipative bosonic bath representing, e.g., lattice or spin collective excitations of the environment.   We apply the method to the Ising-symmetry BFKM with a structureless band and a bath spectral function $\eta(\w)\propto \w^s$. The method is valid for all bath exponents $s$ and all
temperatures $T$. For $0<s<1$, the range of interest in the context of heavy-fermion quantum criticality, an interacting critical point, characterized by hyperscaling of exponents and $\w/T$-scaling, describes a
continuous quantum phase transition between Kondo-screened and localized phases.
For Ohmic dissipation $s=1$, where the model is 
relevant to certain dissipative mesoscopic qubit devices, the transition is found to be Kosterlitz-Thouless-like.   In both regimes the impurity spectral function for the corresponding Anderson model  shows clearly the collapse of the Kondo resonance at the transition. 
 Connection is made to other recent results for the BFKM and the spin-boson model. 
\end{abstract}
\maketitle

    
\section{Introduction}
Quantum impurity models that incorporate dissipative effects of the environment\cite{Leggett:87,Weiss:99} are highly topical.
Applications of such models, which represent the environment by one or more bosonic baths, are numerous and include noisy quantum dots,\cite{LeHur:04,Li:05,Borda:05} decoherence of qubits in quantum computation,\cite{Thorwart:02,Costi:03,Storcz:03,Khveshchenko:04} and charge transfer in biological donor-acceptor systems.\cite{Garg:85,Tornow:06} Furthermore, within dynamical mean-field theory (DMFT) and its extensions,  certain lattice models, of interest, e.g., in relation to heavy-fermion quantum criticality or the interplay of electrons and phonons in strongly correlated electron systems, also reduce to impurity problems of this nature. \cite{Bulla:04a,Lee:06}
In this paper we provide a comprehensive account of the Ising-symmetry Bose-Fermi Kondo model (described further below) obtained via an extension of Wilson's numerical renormalization-group method.\cite{Wilson:75}

The simplest example of a quantum impurity coupling to a bosonic bath is the celebrated spin-boson model (SBM), describing the dynamics of a two-state system coupled to a bath of harmonic oscillators.  For a review see Ref.\ \onlinecite{Leggett:87}.  The essential physics is the
competition between the amplitude for tunelling between the two
spin states $S_z=\pm\frac{1}{2}$ (leading to a ``delocalized phase'') and the effects
of the environment embodied in the coupling, via $S_z$, to the bosonic bath,
which tends to ``localize'' the system in one or other of the spin states.
Not surprisingly, as the generic model for quantum dissipation, the SBM itself finds many applications in physics and chemistry.\cite{Weiss:99}

More general than the SBM are so-called ``Bose-Fermi'' quantum impurity problems involving both fermionic and bosonic baths.
The Bose-Fermi Kondo model (BFKM) was originally discussed in the context of an extended dynamical mean-field theory (EDMFT) treatment of a two-band extended Hubbard model \cite{Smith:99} and describes a local moment coupled not only to a fermionic conduction electron band --- the ``regular'' Kondo model (for a review, see Ref.\ \onlinecite{Hewson:93}) --- but also to a dissipative bosonic bath  specified by a spectral function $B(\w)$ that is typically modeled as $B(\w)\propto \w^s$ at low frequencies.

  Of particular interest is the BFKM with a sub-Ohmic bath $s<1$, which has been studied  within a perturbative (in $\epsilon\equiv 1-s$) renormalization group approach.\cite{Smith:99,Zhu:02,Zarand:02}
 The  coupling of the impurity to the bosonic bath competes with the Kondo effect and leads, for $0<s<1$, to an unstable interacting critical point that describes a continuous quantum phase transition (QPT) between a Kondo-screened phase, in which the impurity moment is quenched by conduction electrons ($\bra S_z\ket = 0$), and a localized phase where $\bra S_z \ket \neq 0$.  Thus, the BFKM provides an example of an \textit{impurity} QPT, encapsulating the competition between local and spatially extended physics.  For a recent review of impurity QPTs, see Ref.\ \onlinecite{Vojta:04a}.

One possible direct realization of the Bose-Fermi Kondo model (BFKM) is a noisy metallic quantum dot: a large micron-scale dot coupled to a reservoir lead and subject to Ohmic dissipation ($s=1$) arising via fluctuations in the gate voltage.  Here, the two impurity ``spin'' states represent the two possible charge states of the dot close to a degeneracy point.  Quantum charge fluctuations on the dot can be supressed by voltage fluctuations,  restoring perfect Coulomb-blockade staircase behavior.\cite{LeHur:04}
The model has also been invoked to study, within a dynamical large-$N$ treatment, the quantum-critical  Kondo effect in a single-electron transistor coupled to both the conduction electrons and spin waves of ferromagnetic leads.\cite{Kirchner:05}

However, to date the BFKM has received most attention in the context of heavy-fermion quantum criticality, arising in an EDMFT treatment of the Kondo lattice model. \cite{Si:01,Si:03,Glossop:06,Zhu:06}  The Kondo lattice model is believed to capture the essential physics underlying heavy-fermion QPTs: the dynamical competition between Kondo screening of $f$-shell moments on each lattice site by conduction electrons, which leads to a paramagnetic  heavy-electron metallic phase; and antiferromagnetic  ordering of those moments due to the Ruderman-Kittel-Kasuya-Yosida (RKKY) interaction. 
In EDMFT, which allows for the nonlocal quantum fluctuations arising from RKKY interactions between local moments, all correlation functions of the Kondo lattice problem can be calculated from a self-consistent Bose-Fermi Kondo model. Here, the bosonic bath embodies the effects, at a given site, of the fluctuating magnetic field generated (via the RKKY interaction) by local moments at other lattice sites.

The above is of considerable current interest in connection with so-called \textit{local} quantum criticality. In this scenario of a heavy-fermion QPT,  destruction of the Kondo effect accompanies divergence of the spatial correlation length at the antiferromagnetic ordering transition.   Such  a local quantum-critical point  may arise within EDMFT --- as a critical impurity solution satisfying the self-consistency --- when magnetic spin fluctuations are two-dimensional. \cite{Si:01,Si:03} The salient non-Fermi liquid features of several materials, including CeCu$_{5.9}$Au$_{0.1}$, which resist the standard spin-density wave description,\cite{Hertz:76} appear to be consistent with a critical point of this nature.  However, 
whether or not a local quantum-critical point (QCP) arises self-consistently under a nonperturbative treatment that considers both paramagnetic and antiferromagnetic phases of the model, is an important question and one that has not been satisfactorily answered by recent quantum Monte Carlo studies. \cite{Grempel:03,Zhu:03,Sun:03}  These studies reach opposing conclusions concerning the nature of the $T=0$ transition, in part due to the inherent limitations of accessing the lowest temperature scales within quantum Monte Carlo.  Similar EDMFT mappings to a BFKM have been applied to the disordered Kondo lattice model  to study interplay of Kondo screening and RKKY interactions in electronic Griffiths phases,\cite{Tanaskovic:04} and to a $t$-$J$ model of a highly incoherent metal close to a Mott transition, \cite{Haule:03} of possible relevance to the pseudgap phase of the cuprates.

Bose-Fermi impurity problems may also arise through a conventional DMFT treatment of lattice models that include  both electron correlation and electron-phonon interactions.  The Hubbard-Holstein model is the simplest such model \cite{Hewson:02,Koller:04a,Koller:04b} and has been discussed in connection with cuprate high-$T_c$ superconductors, heavy-fermion systems, manganites and fulleride superconductivity.  For a recent discussion of Bose-Fermi impurity problems in the context of DMFT, see Ref.\ \onlinecite{Bulla:04a}.

The preceding remarks serve to motivate the study of Bose-Fermi impurity problems as a route to the description of a variety of physically relevant models of strongly-correlated systems.  To what extent the impurity models (or impurity models plus DMFT self-consistency for lattice models) capture the relevant physical behavior can only be assessed if a reliable impurity solver is employed.  Such a technique should be capable of handling static and dynamical properties for all interaction strengths, bosonic bath spectra and temperatures.  This is a demanding requirement, even for purely fermionic impurity models, 
 and many of the standard techniques developed for the latter are here simply inapplicable.
It is natural to turn to Wilson's nonperturbative numerical renormalization-group (NRG)
method,\cite{Wilson:75} which has so far provided a controlled numerical treatment of the
thermodynamics, dynamical response functions, and transport properties of a range of
fermionic impurity models and of lattice models within DMFT.  Recent impurity-model
applications include studies of quantum criticality in the pseudogap Kondo
\cite{Gonzalez-Buxton:98} and Anderson \cite{BGLP:00} models, unconventional physics in
double quantum dots,\cite{Borda:03,Galpin:05,Meden:06,Silva:06} and
non-Fermi liquid
behavior in a frustrated three-impurity Kondo model.\cite{Ingersent:05} Applications to
lattice systems include studies of the Mott-Hubbard metal-insulator transition in the
Hubbard model\cite{Bulla:01} and the paramagnetic phase of the periodic Anderson model
relevant to heavy-fermion systems. \cite{Pruschke:00} For a review, see, e.g., Ref.\
\onlinecite{Hewson:01}.
 The scope of potential applications of the NRG method has recently been widened following a  successful treatment of the spin-boson model via a pure bosonic NRG. \cite{Bulla:03,Bulla:04} The latter has provided a good account of the critical properties of the SBM for both Ohmic and sub-Ohmic bath spectra.

This paper reports a direct study of the Bose-Fermi Kondo model using an NRG method extended to handle  {\it simultaneously} both fermionic and bosonic degrees of freedom. 
We demonstrate that our approach provides a good account of the critical properties of the Ising-symmetry BFKM for both a sub-Ohmic bosonic bath with $s<1$,  as reported briefly in Ref.\ \onlinecite{Glossop:05a}, and for the transition in the Ohmic case $s=1$.  We provide an important confirmation that the Ising-BFKM and SBM belong to the same universality class, as implied by bosonization.
The NRG treatment described is reliable for all bosonic bath exponents and temperatures.  Importantly, it opens the way for a reliable numerical solution of the Kondo lattice model by imposing EDMFT self-consistency on the BFKM solution, as we have discussed elsewhere.\cite{Glossop:06}

The paper is organized as follows. 
After the essential background to the BFKM is given in \sec \ref{secmodel}, an outline of the NRG treatment of the model --- which involves logarithmic discretization of the energy axis, mapping of the Hamiltonian to a chain form, and the iterative numerical solution of the problem at a sequence of decreasing energy scales --- is given in \sec \ref{secnrg}.

Results arising from the approach are presented in two parts, beginning with the case of sub-Ohmic dissipation in Sec.\ \ref{secresults}.  The flow of the NRG transformation and its fixed points is discussed in Sec.\ \ref{secflow}.  An interacting critical fixed point separates a Kondo-screened phase, for weak coupling to the bosonic bath, from a strong-coupling localized phase with no Kondo resonance.  
The form of the phase boundaries, in particular the relationship of the critical coupling to the bosonic bath exponent and the Kondo scale of the pure-fermionic problem, is discussed in \sec \ref{seccrit}.  In \sec \ref{seccross}, for $0<s<1$, we determine the correlation-length exponent that governs the vanishing, at the quantum-critical point, of a low-energy crossover scale.  The response to a local magnetic field applied at the impurity site provides a useful probe of impurity quantum phase transitions; in \sec \ref{secresponse} we discuss the static and dynamical local susceptibility. Distinct behavior obtains in either stable phase and at the QCP.  Critical behavior is manifest as  power laws (with anomalous exponents) in the temperature and frequency dependences, which reflect the fact that the critical physics is scale-free.  We identify an order parameter for the problem, which, for $0<s<1$, vanishes continuously as the transition is approached.  Various critical exponents are extracted and are shown to obey hyperscaling relations.  The latter, together with $\w/T$-scaling at the critical point, are consistent with an interacting fixed point. 
In \sec \ref{secspec} we consider the single-particle spectral function $A(\w)$, in which the collapse of the Kondo resonance at the QCP is directly manifest. The critical properties and scaling behavior  are discussed in detail.  

The nature of the transition changes qualitatively at $s=1$, as reflected in the divergence of the correlation length exponent as $s\ra 1^-$.   
We consider the Ohmic case ($s=1$) in \sec \ref{s1section}, demonstrating that the transition is Kosterlitz-Thouless-like, with characteristic vanishing of the renormalized Kondo scale and a jump in the order parameter at the critical coupling.  The spectral function and its scaling properties are discussed and compared to their sub-Ohmic counterparts.  We also briefly examine our results in the context of the noisy quantum box.\cite{LeHur:04,Li:05,Borda:05}

The paper concludes with a brief summary and outlook for future work.

\seceq
\section{BACKGROUND}
In the remainder of this paper we focus on the Bose-Fermi Kondo model with Ising anisotropy in the bosonic couplings.  This is the case most relevant to the EDMFT treatment of CeCu$_{6-x}$Au$_x$  and to the noisy quantum dot system.  We treat the model by extending the NRG method to encompass Bose-Fermi impurity problems.  Though described in Sec.\ \ref{secnrg} for the Ising-BFKM, the approach is straightforwardly generalized to the $xy$ and isotropic problems.  Results for the latter cases will be discussed elsewhere.  In the following subsection we define the model to be studied and identify key quantities, before going on to discuss the connection to the spin-boson model in Sec.\ \ref{mapping}. We set $\hbar=g\mu_{\mbox{\ssz{B}}}=k_{\mbox{\ssz{B}}}=1$.

\label{secmodel}
\subsection{Ising-anisotropic Bose-Fermi Kondo model}
The Hamiltonian for the Ising-symmetry Bose-Fermi Kondo model is given by \be
\hat{\cH}=\hat{\cH}_F+\hat{\cH}_B + \hat{\cH}_{\mbox{\ssz{int}}}, \label{hamdef}\ee where
\be \hat{\cH}_F = \sum_{\kk, \sigma}
\epsilon_{\kk}c^{\dagger}_{\kk\sigma}c_{\kk\sigma} \label{hfermi}\ee and \be
\hat{\cH}_B = \sum_{\qq}\w_{\qq}\phi^{\dagger}_{\qq}\phi_{\qq} \ee
describe, respectively, the conduction band and a dissipative 
bosonic bath.  Both are taken to be noninteracting and
characterized by the densities of states \be
\rho(\epsilon)=\sum_{\kk}\delta(\epsilon-\epsilon_{\kk}) \label{rhofermi}\ee and
\be \eta(\w)=\sum_{\qq}\delta(\w-\w_{\qq}).\label{bosedos}\ee The interaction of
the impurity with the baths is contained in \be
\hat{\cH}_{\mbox{\ssz{int}}}=\mbox{$\frac{1}{2}$}J_0\bi{S} \cdot
\sum_{\kk, \kk', \sigma,
\sigma'}c_{\kk\sigma}^{\dagger}\mbox{\boldmath$\sigma$}_{\sigma\sigma'}c_{\kk'\sigma'}
+ S_z \sum_{\qq} g_{\qq}(\phi_{\qq} + \phi^{\dagger}_{-\qq}).
\label{hint} \ee The first term describes the interaction of a
spin-$\frac{1}{2}$ local moment with the on-site spin of the
fermionic band via the usual Kondo coupling $J_0$; the second
term describes its coupling, of magnitude $g_{\qq}$, to the bosonic
bath with oscillators $\{\phi_{\qq}\}$.

We assume a featureless metallic host conduction band of width $2D$ described by
a density of states
    \be
    \rho(\epsilon)=\rho_0 \Theta(D-|\epsilon|)
    \label{fermidos}
    \ee
with $\Theta(x)$ the Heaviside step function.   The bosonic bath is entirely determined by the spectral function
\be
B(\w)=\pi \sum_{\qq} g^2_{\qq}\delta(\w-\w_{\qq}),
\label{bathspec}
\ee
which we assume vanishes outside the range $0<\w<\w_0$.
Note that the bath density of states $\eta(\w)$ and coupling $g_{\qq}$ do not require
separate specification.  Writing $x=\epsilon/D$ and $y=\w/\w_0$, the Hamiltonian is conveniently written in the following one-dimensional form: 
\alpheqn
\be \hat{H}_F + \hat{H}_B = D\int_{-1}^{+1}
\mbox{d}\ve\  \ve~
c_{\ve\sigma}^{\dagger}c_{\ve\sigma} + \w_0\int_0^{1}
\mbox{d}\vw\ \vw~\bphi_{\vw}^{\dagger}\bphi_{\vw}, \ee \be
\hat{H}_{\mbox{\ssz{int}}} = D\rho_0J_0\bi{S}\cdot \sum_{\sigma\sigma'}
f_{0\sigma}^{\dagger}\mbox{\boldmath$\sigma$}_{\sigma\sigma'}f_{0\sigma'}
+ \w_0\sqrt{\frac{B_0 \cb^2 }{\pi}}{S_z}({b}_0+{b}_0^{\dagger}),
\label{1dhint}
     \ee
\reseteqn
extracting a dimensionless ``dissipation strength'' $B_{0}$ from $B(\w)$.
 The dimensionless operators
 \be f_{0\sigma}=\frac{1}{\sqrt{2}}\int_{-1}^{+1} \mbox{d}\ve\
c_{\ve\sigma} \ee
    and
 \be {b}_0=\cb^{-1}\int_0^{1}\mbox{d}\vw
\ W(\vw) \phi_{\vw},
\label{b0def}
    \ee
the latter defined by a  weighting function
\be
W(\vw)=\sqrt{\frac{B(\w_0 \vw)}{B_0\w_0}}
\ee
and a normalization factor
  \be
\cb^2=\int_0^{1}\mbox{d}\vw\ W^2(\vw),
    \ee
respectively annihilate the unique combination of fermions and
bosons that couples to the impurity.    The operators $c_{\ve\sigma}$ and $\phi_{\vw}$ obey the commutation relations $\{c_{\ve\sigma}^{\dagger},c_{\ve'\sigma'}\}=\delta(\ve-\ve')\delta_{\sigma\sigma'}$ and $[\phi_{\vw}^{\dagger},\phi_{\vw'}]=\delta(\vw-\vw')$ respectively, while $\{f_{0\sigma}^{\dagger},f_{0\sigma'}\}=\delta_{\sigma\sigma'}$ and $[b_0,b_0^{\dagger}]=1$.

Of principal interest is the low-temperature physics of the model, determined by the form of the bath spectrum at low energies.  We take $B(\w)$ to be given by
\be
B(\w)=\left\{ \begin{array}{ll}
B_{0~}\w_0^{1-s}\w^s& \mbox{\ \ \ for\ } 0<\w<\w_0,\\
0 & \mbox{\ \ \ otherwise,}
\end{array} \right.
\label{bosebath}
\ee
with $s>-1$ necessarily.   The bulk of the paper concerns the sub-Ohmic regime $s<1$; in particular we focus on $0<s<1$ where the key physics is governed by an interacting quantum-critical point.  However, we also study the Ohmic case ($s=1$) in \sec \ref{s1section}.

To connect with
previous work on the BFKM, we will assume that $g_{\qq}$ is
$\qq$-independent ($g_{\qq}\equiv g_0$) and that the bath spectral function is of form
\be
\eta(\w)=\left\{ \begin{array}{ll}
(K_0^2/\pi)~\w_0^{1-s}~\w^s & \mbox{\ \ \ for\ }0<\w<\w_0\\
0 & \mbox{\ \ \ otherwise,}
\end{array} \right.
\label{etaofw}
\ee
and hence, from Eqs.\ (\ref{bathspec}) and (\ref{bosedos}), $B_0\equiv(K_0g_0)^2$.

\subsection{Connection to the spin-boson model}
\label{mapping}
The spin-boson model is the simplest model of quantum dissipation, \cite{Leggett:87} describing a two-level system coupled to a bosonic bath:
\be
\hat{\cH}_{SB}=\hat{\cH}_{B}-{\Delta}S_x+\epsilon S_z+S_z\sum_{\qq} \lambda_{\qq}(\phi_{\qq} + \phi^{\dagger}_{-\qq}).
\label{hamsbm}
\ee
Here, $\Delta$ is the amplitude for tunneling between the ``spin'' states $S_z=\pm\frac{1}{2}$, and $\epsilon$ represents an additional bias.  The coupling to the environment is characterized by the spectral function $J(\w)=\pi\sum_{\qq}\lambda_{\qq}^2\delta(\w-\w_{\qq})$.   It is well known\cite{Leggett:87} that there exists a correspondence between the Ohmic spin-boson model and the anisotropic Kondo model (AKM), described by a Hamiltonian $\hat{H}=\hat{H}_F+\hat{H}_{\mbox{\ssz{int}}}$ with $\hat{H}_F$ given by Eqs.\ (\ref{hfermi}), (\ref{rhofermi}) and (\ref{fermidos}), and
\bea
\hat{H}_{\mbox{\ssz{int}}}&=&\frac{J_{\perp}}{2}\sum_{\kk,\kk'}(c_{\kk\up}^{\dagger}c_{\kk'\down}S^-
+ c_{\kk\down}^{\dagger}c_{\kk'\up}S^+) \nonumber \\ 
&+&\frac{J_{z}}{2}\sum_{\kk,\kk'}(c_{\kk\up}^{\dagger}c_{\kk'\up}- c_{\kk\down}^{\dagger}c_{\kk'\down})S^z +hS^z.
\label{akmham}
\eea
   Specifically, following  bosonization of the AKM (see, e.g., Ref.\ \onlinecite{Costi:99}),
\bea
\epsilon&=&g\mu_Bh,\\
\frac{\Delta}{\w_c}&=&\rho_0J_{\perp},\\
J(\w)&=&2\pi\alpha \w \ \ \ \mbox{for} \ \w<\w_c,
\eea
where the dimensionless dissipation strength $\alpha=[1-(2/\pi)\tan^{-1}(\pi\rho_0 J_{z}/4)]^2$ and $\w_c=2D$.

 The Ohmic case of the SBM is well understood.   The essential physics is a Kosterlitz-Thouless quantum phase transition between a delocalized phase for $\alpha<\alpha_c$ --- in which the effect of dissipation is to renormalize the tuneling amplitude --- and a localized phase for $\alpha>\alpha_c$, where tunneling between the spin states is absent.  For $\Delta\ra 0$, $\alpha_c=1$.
Thus, the Kondo scale $T_{K}$ of the AKM can be interpreted as a renormalization of the bare tunneling amplitude $\rho_0J_{\perp}$ due to frictional effects of the electronic environment (particle-hole excitations) controlled by $\rho_0J_{z}$.  The antiferromagnetic phase of the AKM  corresponds to the delocalized phase of the SBM, while the ferromagnetic sector  corresponds to the localized phase.  For the regular isotropic Kondo model, $J_{z}=J_{\perp}\equiv J_0$, the corresponding SBM dissipation strength takes the value $\alpha=1^-$ for $\rho_0J_0\ll 1$.

The SBM with sub-Ohmic dissipation has only recently been studied in detail, using perturbative \cite{Vojta:04} and numerical renormalization group \cite{Bulla:03} methods.  For bath exponents $0<s<1$ these studies establish the existence of an interacting quantum-critical point
 describing a continuous QPT between delocalized and localized phases, with $\alpha_c\propto \Delta^{1-s}$ for $\Delta \ll \w_c$.  

Recent studies of the Ising-symmetry Bose-Fermi Kondo model with Ohmic dissipation\cite{Li:05,Borda:05} have exploited the connection between the AKM and Ohmic SBM.   For example, the essential results of Ref.\ \onlinecite{Bulla:04} have been discussed in the context of the Ohmic BFKM, \cite{Li:05} of relevance to certain dissipative mesoscopic qubit devices.  More  generally, the Ising-symmetry BFKM can be mapped to a spin-boson model whose spectral function $J(\w)$ comprises an Ohmic contribution from the bosonized conduction band  and a contribution $B(\w)\propto\w^s$ from the BFKM bath. The resulting $J(\w)$ is dominated by $B(\w)$ at low frequencies and thus it is to be expected that the Ising symmetry BFKM and SBM have identical universal critical properties for $s\le 1$.  However, the presence of the Ohmic contribution acts to renormalize the bare tunneling amplitude entering the model, such that the sub-Ohmic BFKM corresponds to a SBM, Eq.\ (\ref{hamsbm}), with $J(\w)\propto \w^s$ for $\w \ll \w_c$ and the bare Kondo scale $\tkn$ (not $J$) the appropriate tunneling amplitude.  This is borne out by the numerical results of Sec.\ \ref{secresults}.

 \seceq
\section{NRG treatment}
\label{secnrg}

In  Wilson's original NRG treatment of the
Kondo model\cite{Wilson:75} [as described by
$\hat{H}=\hat{H}_F+\hat{H}_{\mbox{\ssz{int}}}$, Eqs.\ (\ref{hfermi}), (\ref{hint}) with $g_{\qq}=0$, and (\ref{fermidos})] the
conduction band continuum of energies $\epsilon$ is sampled at the discrete set of energies  $\epsilon=\pm D\Lambda^{-m}$ for $m=0, 1, 2, ...$, where $\Lambda>1$
parametrizes the discretization.  
By a mapping of $\hat{H}_F$ to a tight-binding Hamiltonian
using the Lanczos procedure, the Kondo Hamiltonian is cast in the form of a semi-infinite 
fermionic chain with the impurity coupled only to the end site $n=0$.
Owing to the discretization, the tight-binding coefficients decay with site number as $D\Lambda^{-n}$ for large $n$, which feature allows for an iterative solution to the problem by numerically diagonalizing the Hamiltonian for progressively longer finite chains.  The  procedure amounts to including excitations on an exponentially smaller energy scale at each step, with the full Hamiltonian being recovered as the limit
\be
\hat{H}=\lim_{N\ra\infty}\alpha\Lambda^{-{N}/{2}}D{H}_N.
\label{Hlimit}
\ee
Here, $\hat{H}_N$ describes an $(N+1)$-site (fermionic) chain, rescaled so that the lowest energy is of order unity and a meaningful comparison between the $\{\hat{H}_N\}$ can be made. The constant $\alpha$ is conventional and given by $\alpha=\frac{1}{2}\Lambda^{\frac{1}{2}}(1+\Lambda^{-1})$. 

The renormalization-group transformation $\hat{R}$, defined by ${H}_N=\hat{R}{H}_{N-1}$, eventually reaches a 
scale-invariant fixed point of $\hat{R}^2$ that determines the low-temperature properties of the system.   At a fixed point, the low-lying eigenstates and matrix elements of any observable operator are identical for iterations $N$ and $N-2$.  That the fixed points are of the transformation $\hat{R}^2$ and not $\hat{R}$ reflects the fundamental inequivalence of fermionic chains containing odd and even numbers of sites, as discussed, e.g., in Ref.\ \onlinecite{Krishna-murthy:80}.

In the following we outline the transformation of the BFKM, Eqs.\ (\ref{hamdef}) and 
(2.9), to the form required for an NRG treatment. Including a coupling between the impurity and a bosonic bath introduces two additional considerations.  First, and most obviously, the bosonic Hilbert space must be truncated from the outset, introducing a bosonic truncation parameter.  The effects of this truncation on the critical properties of interest must be systematically studied.  Second, the iterative procedure employed must preserve the spirit of the NRG approach, taking into account all excitations of the same energy scale at the same iteration.  We turn to a discussion of these issues in the following section.

\subsection{Discretization and mapping}
 The bosonic bath [conduction band] continuum of energies
 $0\le\omega\le \w_0$ [$-D \le \epsilon\le D$]  is
replaced by a discrete set: $\w_0\Lambda^{-m}$ [$\pm D\Lambda^{-m}$] for
integer $m\ge 0$.  
The main approximation of the NRG is to select just one bosonic [fermionic] state from each energy interval.  We retain the single state that couples directly to the impurity  such that $b_0$ in Eq.\ (\ref{b0def}) is given exactly by
\be 
b_0=\cb^{-1}\sum_m\cb_m~ \phi_m
\ee 
with
\be
\phi_m=\cb_m^{-1}\int_{\Lambda^{-(m+1)}}^{\Lambda^{-m}}\mbox{d}\vw~W(\vw)~\phi_{\vw}
\ee
and
  \be \cb_m^2=\int_{\Lambda^{-(m+1)}}^{\Lambda^{-m}}\mbox{d}\vw~W^2(\vw).
    \ee
For the fermions,
\be
f_{0\sigma}=\frac{1}{\sqrt{2}}\sum_m (c_{m\sigma}^{>}+c_{m\sigma}^{<}).
\ee
The operators $c_{m\sigma}^>$  and $c_{m\sigma}^<$ are defined for the $m$th positive and negative energy intervals respectively and are given by
\bea
c_{m\sigma}^>&=&\int_{\Lambda^{-(m+1)}}^{\Lambda^{-m}}\mbox{d}\ve\ c_{\ve\sigma},\\
c_{m\sigma}^<&=&\int_{-\Lambda^{-m}}^{-\Lambda^{-(m+1)}}\mbox{d}\ve\ c_{\ve\sigma}.
\eea
The terms $\hat{H}_B$ and $\hat{H}_F$ are then approximately written as
\be
\hat{H}_B\simeq  \w_0 \sum_m^{\infty}\w_m ~\phi_m^{\dagger}\phi_m,
\label{disbath}
\ee
\be
\hat{H}_F\simeq D \sum_m^{\infty}(\epsilon_m^> c_{m\sigma}^{>\dagger}c_{m\sigma}^>
+ \epsilon_m^< c_{m\sigma}^{<\dagger}c_{m\sigma}^<),
\label{disband}
\ee
where
\be
\w_m=\cb_m^{-2}\int_{\Lambda^{-(m+1)}}^{\Lambda^{-m}}\mbox{d}\vw~W(\vw)\vw,
\label{omegam}
\ee 
\be
\epsilon_m^{\gtrless}=\pm\frac{1}{2}\left(\Lambda^{-m}+ \Lambda^{-(m+1)}\right),
\ee
and the approximation is seen to amount to selecting the state of weighted mean energy from each interval. The infinite number of states neglected in each interval are orthogonal to the state retained and couple only indirectly to the impurity, which coupling vanishes as $\Lambda\ra 1$.  As is standard in the NRG approach, we neglect this coupling for $\Lambda>1$ and drop the irrelevant decoupled constant term that thereby results.  The discretization error incurred 
 vanishes as $\Lambda \ra 1$, which fact alone favors
choosing $\Lambda$ close to unity.  One manifestation of discretization error
 is a modification of the effective coupling between impurity and bosonic bath [conduction band].  Essentially exact results, if desired, can be obtained by performing a $\Lambda\ra 1^+$ extrapolation from $\Lambda>1$ results.
Fortunately, critical exponents are often found to be extremely insensitive to the chosen $\Lambda$.  Larger $\Lambda$ (e.g., $\Lambda = 9$) calculations, which retain fewer states and thus require less computational effort, can then be used to extract the critical properties of the model.  However, the weak $\Lambda$-dependence of critical exponents cannot of course be assumed at the outset.

\begin{figure}
\begin{center} \epsfig{file=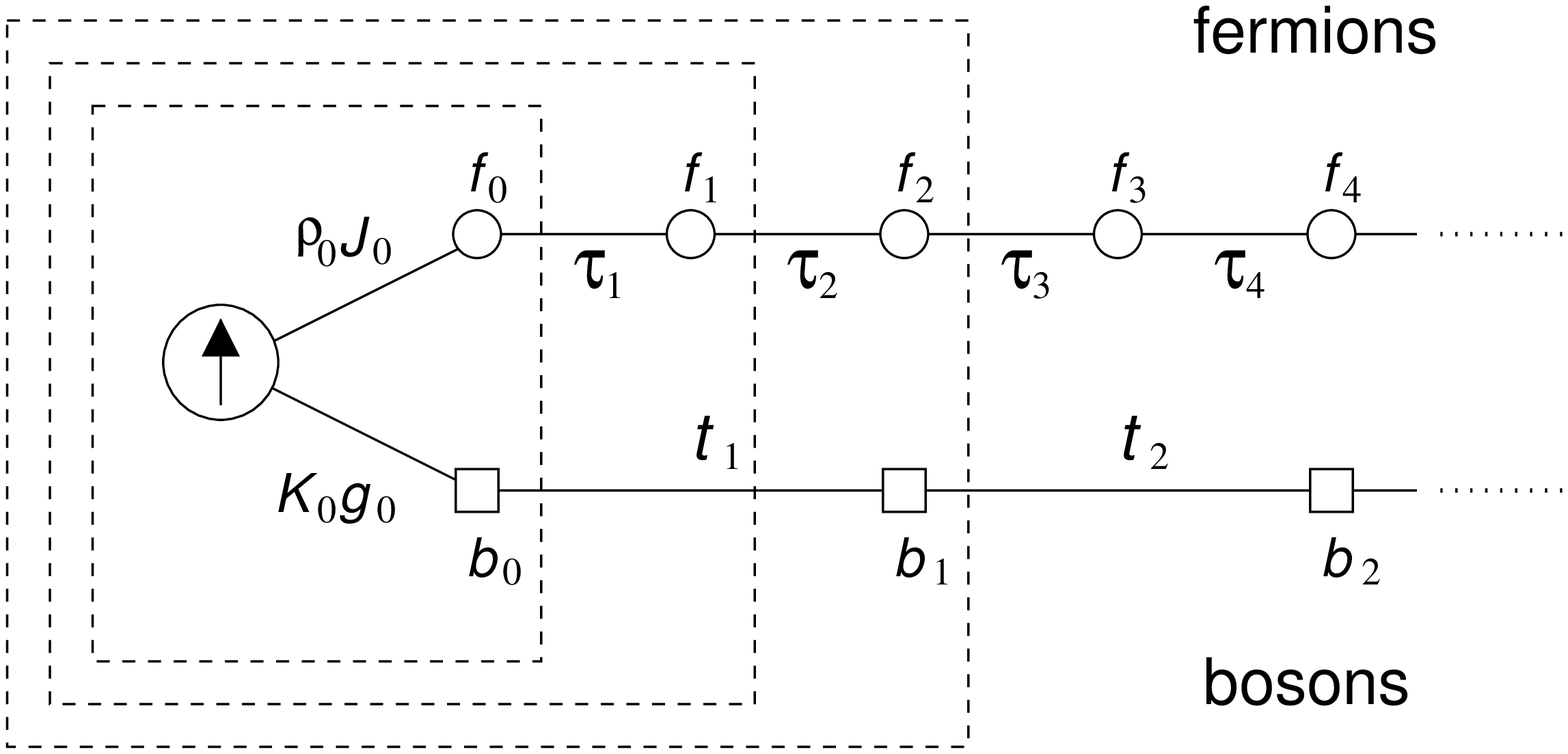, width=8cm} \end{center}
\caption{Schematic of the chain form of the BFKM Hamiltonian.  Fermionic sites of the tight-binding chain are shown as small circles, while bosonic sites are shown as squares.  Only the end site of each chain couples directly to the impurity spin.  The coupling between the
fermionic sites decays as $\tau_n\sim\Lambda^{-n/2}$ for large $n$, while the bosonic couplings fall off as $t_n\sim\Lambda^{-n}$.  The dashed boxes enclose, from innermost to outermost, the sites included in iterations 0, 1, and 2, respectively.} \label{chainfig}
\end{figure}

Finally, Eqs.\ (\ref{disbath}) and (\ref{disband}) are mapped exactly onto  tight-binding Hamiltonians via a discretized Lanczos procedure.\cite{Lanczos:50}  Defining 
\be
b_n=\sum_m u_{nm} \phi_m,
\ee
\be
f_{n\sigma}=\sum_m (v_{nm}^{>} c_{m\sigma}^{>}+v_{nm}^{<}c_{m\sigma}^{<}),
\ee
where $u_{0m}=\cb_m/\cb$ and $v_{0m}^{\gtrless}=1/\sqrt{2}$, such that $[b_n,b_{n'}^{\dagger}]=\delta_{n,n'}$ and $\{f_{n\sigma},f_{n',\sigma'}^{\dagger}\}=\delta_{n,n'}\delta_{\sigma,\sigma'}$, we obtain
\be
\hat{H}_B=\w_0 \sum_n \left[ e_n b_n^{\dagger}b_n + t_n(b_n^{\dagger}b_{n-1}
+b^{\dagger}_{n-1}b_n) \right],
\label{hbchain}
\ee
\be
\hat{H}_F=D \sum_{n,\sigma} \left[ \epsilon_n f_{n\sigma}^{\dagger}f_{n\sigma}
+\tau_n(f_{n\sigma}^{\dagger}f_{n-1,\sigma}+f_{n-1,
\sigma}^{\dagger}f_{{n}\sigma}) \right].
\label{hfchain}
\ee
The tight-binding coefficients $\epsilon_n$, $\tau_n$, $e_n$ and $t_n$ ($n = 0, 1, 2, ...$) encode all information about the conduction band and bosonic bath and are determined numerically, for given $B(\w)$ and $\rho(\w)$, via Lanczos recursion relations.  For example, $e_n$ and $t_n$ follow from numerically iterating
\bea
\sum_m u_{nm}^2\w_m &=& e_n, \nonumber\\
u_{nm}(\w_m-e_n)-t_n u_{n-1,m}&=&t_{n+1}u_{n+1,m},\ \nonumber \\
\sum_m u_{n+1,m}^2 &=& 1,
\eea
with $\wm$ given by Eq.\ (\ref{omegam}).
For a symmetric fermionic band density of states $\rho(\epsilon)=\rho(-\epsilon)$,  $\epsilon_n=0$ for all $n$. For the specific case Eq.\ (\ref{fermidos}) considered in the following, we use the closed-form algebraic expression for $\tau_n$ as derived by Wilson.\cite{Wilson:75}  
The transformed Hamiltonian to be solved, $\hat{H}=\hat{H}_F+\hat{H}_B+\hat{H}_{\mbox{\ssz{int}}}$ [Eqs.\ (2.9), (\ref{hbchain}) and (\ref{hfchain})], describes two semi-infinite chains each coupled at the end site only to a spin-$\frac{1}{2}$ magnetic moment.  A schematic of the chain form of the Hamiltonian is depicted in Fig.~(\ref{chainfig}).   

\subsection{Iterative solution}
\label{secitsol}
The  on-site energies $\epsilon_n$ and hopping coefficients $\tau_n$ for the fermionic tight-binding chain  decrease like $\Lambda^{-n/2}$ for large chain length $n$.
The hopping coefficients for the bosonic chain, $e_n$ and $t_n$, drop off faster --- as $\Lambda^{-n}$ for large $n$ --- arising because the spectral density $B(\w)$ is nonzero for $\w>0$ only.  
The latter fact has been discussed in the context of a pure \textit{bosonic} NRG approach to the spin-boson model, \cite{Bulla:03,Bulla:04}  and there allows a convergent approximation using a single bosonic chain of finite length.
While the BFKM mapping is essentially a combination of the pure fermionic and pure bosonic approaches, the
iterative procedure described below is an important novel feature of our
approach, and one  that does not {\it a priori} guarantee a satisfactory description of the quantum-critical physics of the model.

In employing the NRG for the BFKM we must ensure that at each  step we treat only fermions and bosons of the same energy scale.
As for the pure fermionic case, we take  $D$ as the energy unit and write the full Hamiltonian  as the limit Eq.\ (\ref{Hlimit}), where here $N$ refers to the highest labelled fermionic site or iteration number.  Crucially, the following recursion relation,  relating ${H}_N$ to ${H}_{N-1}(M-1)$, adds site $M$ to the bosonic chain only if  its tight-binding coefficients are comparable to the energy scale ($\Lambda^{-N/2}$) set by the fermionic site $N$  added at the current iteration, $N$.  Thus, 
\bea
{H}_N&=&\Lambda^{\frac{1}{2}}{H}_{N-1}(M-1) + \alpha^{-1}\Lambda^{N/2} \sum_{\sigma}\epsilon_Nf_{N\sigma}^{\dagger}f_{N\sigma}\nonumber \\
 &+& \alpha^{-1}\Lambda^{N/2} \sum_{\sigma} \tau_N(f_{N\sigma}^{\dagger}f_{N-1,\sigma} + \mbox{H.c.})\nonumber \\
&+&\wot \alpha^{-1}\Lambda^{N/2}\Theta(\xi_{M}-c\Lambda^{-N/2})\left[e_Mb_M^{\dagger}b_M \right. \nonumber \\
&+& \left. t_M(b_M^{\dagger}b_{M-1}+ \mbox{H.c.})\right]
\label{hamn}
\eea
where $\wot=\w_0/D$, $\xi_{M}=\max(e_M,t_M)$ and $c$ is a threshold parameter that we typically assign the value $c=\Lambda^{-1/4}$. The Hamiltonian for iteration $N=0$, containing the  couplings between the impurity and the end site of each chain, is given by
\bea
\alpha{H}_0&=&\sum_{\sigma \sigma'}{\rho_0 J_0}\bi{S}\cdot f_{0\sigma}^{\dagger}\bi{\sigma}_{\sigma\sigma'}f_{0\sigma'}+\sum_{\sigma}\epsilon_0f_{0\sigma}^{\dagger}f_{0\sigma}\nonumber\\
&+&\wot \sqrt{\frac{ B_0 \cb^2}{\pi}}S_z(b_0^{\dagger}+b_0)+\wot e_0b_0^{\dagger}b_0,
\eea
with natural dimensionless parameters  $\rho_0J_0$, $B_0\equiv(K_0g_0)^2$ and $\wot=\w_0/D$ to be varied for a given exponent $s$ in the bath spectral function $B(\w)$ [Eq.\ (\ref{bathspec})].

Since both $e_n$ and $t_n$ decay as $\Lambda^{-n}$ for large chain length $n$, the iterative procedure Eq.\ (\ref{hamn}) rapidly amounts to extending the bosonic chain at every second iteration (i.e., for every second site added to the fermionic chain).  Thus without any change in the essential physical properties, it is possible, and proves convenient,  to modify Eq.\ (\ref{hamn}) to force the extension of the bosonic chain to occur only at a given iteration parity.

We also note in passing that an alternative NRG scheme for the BFKM might extend both chains at each iteration but employ a different discretization parameter $\Lambda_B$ for bosonic energies related to that for fermionic energies $\Lambda_F$ by $\Lambda_{F}=\Lambda_B^2$ (Ref.\ \onlinecite{Bulla:priv}).  Though such a scheme has not yet been implemented, we believe it does not confer any significant advantages over the approach taken in this work.

The eigenstates $\{|E\ket_{N-1}\}$ of ${H}_{N-1}$ are used to construct the basis states for iteration $N$:
\be
|l,E\ket_N= |E\ket_{N-1}\otimes |l\ket_N.
\label{basis}
\ee
where $\{|l\ket_N\}$ are the states of the added site(s).  If  the fermionic chain alone is extended then $|l\ket_N\equiv|i\ket_N$, one of the four possible states of the added fermionic site.
If both a  fermionic and a bosonic site are added, then $|l\ket_N=|i\ket_N\otimes |n_{M}\ket$ where$\{|n_{M}\ket\}$ is an eigenstate of $b_M^{\dagger}b_M$. In contrast to fermionic sites, the number of bosons per site of the bosonic chain is in principle unlimited.   In practice, the bosonic Hilbert space must be truncated, restricting the maximum number of bosons per site of the bosonic chain to a finite number $N_b$, typically $N_b\le 16$ and thus including only the lowest $N_b+1$ eigenstates of $b_M^{\dagger}b_M$.  We have systematically studied the effects of this truncation on various properties of the BFKM.  In particular, we find that universal critical properties rapidly converge with increasing $N_b$  --- typically $8<N_b<16$ suffices --- providing a very satisfactory description of the QCP for the model.   This is discussed further in Sec.\ \ref{secresults}. 

It follows from Eqs.\ (\ref{hamn}) and (\ref{basis}) that  the matrix elements of ${H}_N$, $_N\bra l',E'|{H}_N|l,E\ket_N$, can be evaluated in terms of the eigenstates and eigenenergies of ${H}_{N-1}$ and the matrix elements $_N\bra i'|f_N^{\dagger}|i\ket_N$ and, if appropriate, $_M\bra n'|b_M^{\dagger}|n\ket_M$. 
The numerical effort required to diagonalize ${H}_N$ is considerably reduced by taking advantage of various symmetries of the Hamiltonian, as discussed, e.g.,  in Refs.\ \onlinecite{Krishna-murthy:80} and \onlinecite{Gonzalez-Buxton:98}.  With ${H}_N$ diagonalized, the matrix elements of ${H}_{N+1}$ are formed in the basis Eq.\ (\ref{basis}) with $N\ra N+1$ and the procedure is repeated to obtain the many-body spectrum on progressively lower energy scales.

The need for an iterative approach arises because (even after bosonic truncation) the size of the basis rapidly becomes unmanageable.  For the regular Kondo model it grows by a factor of $4$ at each iteration, reflecting the possible occupancies of the added fermionic site.   Even in
that relatively simple application, finite computational power demands that the basis  be truncated after only a few iterations, saving
only the lowest $N_s$ many-body eigenstates of $H_{N-1}$ to be used to construct the basis in which to diagonalize $H_{N}$.  The problem is of course even more acute in the BFKM: for each iteration that a bosonic site is added to the chain the basis grows by a factor of $4(N_b+1)$.

By retaining only the lowest $N_s$ eigenstates from one iteration to the next, the NRG method gives a sequence of truncated Hamiltonians from which thermodynamic [dynamical] properties may be calculated.  For each iteration $N$, the error due to truncation is minimized in a temperature [frequency] window around a characteristic energy scale 
\be
T_N=\alpha\Lambda^{-N/2}/\bar{\beta},
\label{tempdef}
\ee
where $\bar{\beta}<1$ is a dimensionless parameter \cite{Gonzalez-Buxton:98} that generally decreases with increasing $N_s$.  Properties at temperature [frequency] scales smaller than $T_N$ will be more accurately calculated at subsequent iterations, while those at higher scales have been more accurately calculated at a previous iteration.  See, e.g., Refs.\ \onlinecite{Krishna-murthy:80} and \onlinecite{Costi:94} for further discussion.

In the following we turn to results arising from the above approach, beginning with the case of sub-Ohmic dissipation.

\begin{figure}
\begin{center}
\epsfig{file=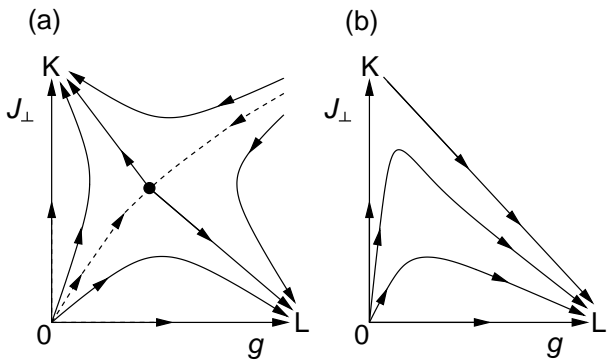, width=8cm}
\end{center}
\caption{Schematic of the NRG flow of effective couplings for  (a) $0<s<1$ and (b) $-1<s\le 0$.   Trajectories describe the flow with increasing $N$ (or decreasing $T_N$) of the effective  $J_{\perp}$ and $g$ that enter ${H}_N$.   For $0<s<1$ an unstable critical fixed point (solid circle) lies on the separatrix (dashed line) that specifies the phase boundary between the stable Kondo (K) and localized (L) phases in the $J_{\perp}$-$g$ plane.  The Kondo fixed point corresponds to an effective $J_{\perp}=J_z=\infty$ and $g=0$, while the localized fixed point has $J_{\perp}=0$,  $J_z$ finite, and $g=\infty$.  For $-1<s\le 0$ the localized fixed point is reached for all $g\ne 0$. See text for further discussion.}
\label{flow}
\end{figure}

\seceq
\section{Results: Sub-Ohmic dissipation}
\label{secresults}
We now consider results for the Bose-Fermi Kondo model with sub-Ohmic dissipation, i.e.,\ $s<1$ in Eq.\ (\ref{bosebath}).  The Ohmic case $s=1$ is discussed separately in Sec.\ \ref{s1section}. For $0<s<1$ the model exhibits a continuous impurity quantum phase transition between Kondo-screened and unscreened-local-moment  phases.  After discussing the NRG flows, fixed points and phase boundaries, we demonstrate that a low-energy scale $T_*$, characterizing the crossover from critical to stable fixed point behavior, vanishes in a power-law fashion at the quantum-critical point.  Local magnetic properties  considered in Sec.\ \ref{secresponse}  show distinct behaviors in either phase and at the QCP itself, from which various critical exponents can be extracted.  The interacting nature of the QCP is evident from hyperscaling of the criticial exponents, including $\w/T$-scaling behavior in the dynamics.  We conclude the discussion of the sub-Ohmic case by considering the local spectrum of single-particle excitations for the related Bose-Fermi Anderson model.  The destruction of the central Kondo resonance at the QCP is clearly observed.  
  
For notational compactness  we write henceforth $\rho_0J_0\equiv J$ and $K_0g_0\equiv g$.  We also set the bosonic bath cut-off $\wot=1$ in our calculations,\cite{cutoff}  although we emphasize that our approach is in no way limited to this particular choice.  All energies, frequencies, and temperatures are expressed as multiples of $D=1$.
For a given bath exponent $s$, we  typically work at fixed $J$ and vary $g$ to tune the system through a quantum phase transition.  However, in calculations of the single-particle spectrum, we specify $B_0\equiv g^2$. Critical properties are found to be insensitive to the NRG discretization parameter $\Lambda$ and adequately converged for bosonic truncation parameter $8<N_b<12$.  Unless stated otherwise, results presented were obtained using $\Lambda=9$, $N_b=8$ and $N_s=500$.

\subsection{NRG flow and fixed points}
\label{secflow}
Figure \ref{flow}(a) shows a schematic of the NRG flow of effective couplings found for $0<s<1$.  There are two stable fixed points of the flow: the familiar Kondo fixed point (K), where the impurity spin is screened by conduction electrons and the single-particle spectrum contains a Kondo resonance (see Sec.\ \ref{secspec} ); and a ``localized'' fixed point (L), where the impurity dynamics are controlled by coupling to the dissipative bath. A third fixed point --- the unstable, critical fixed point (where Kondo screening is critical) --- lies on the separatrix $g=g_c(J)$ which specifies the phase boundary in the $J$-$g$ plane (see Sec.\ \ref{seccrit}).

\begin{figure*}
\begin{center}
\epsfig{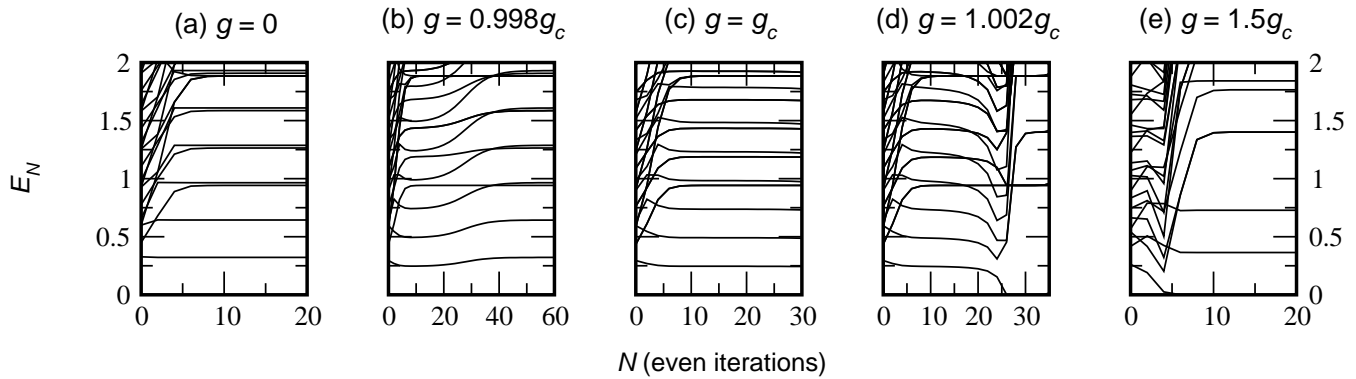}
\end{center}
\caption{Low-lying energy levels of ${H}_N$ as a function of even iteration number $N$, for bosonic bath exponent $s=0.2$, Kondo coupling $J=0.5$ and the values of $g$ shown, with $g_c\simeq 0.567$.  The bosonic chain was extended for even $N$ only, as discussed in Sec.\ \protect \ref{secitsol}.  (a) Zero coupling to the bosonic bath.  (b) For $g$ close to $g_c$ in the Kondo phase, the quantum-critical fixed-point regime [shown in (c)] is accessed for intermediate $N$ (here $10\lesssim N \lesssim 20$) before the levels flow to the Kondo-phase fixed-point structure (for $N\gtrsim 40$).  (d) and (e) In the localized phase, the levels cross over from the critical fixed-point [$10\lesssim N \lesssim 20$ in (d)] to the localized fixed-point structure observed for large $N$.  The crossover in the flows defines a low-energy scale $T_*$ that vanishes as $g\ra g_c$.
 }
\label{levelflowk}
\end{figure*}

\begin{table}[b]
\caption{\label{tab:table1} Low-lying eigenstates of the even-$N$ fixed-point Hamiltonian ${H}^*_K$ obtained for $s=0.2$. The $\{\ff_i\}$ denote the limiting eigenvalues of $H_{F,N}^{(1)}$ [Eq.\ (\protect\ref{freefermi})] corresponding to the regular Kondo model fixed-point spectrum, while $\{\fb_i\}$ are the limiting eigenvalues for a chain of free bosons.}
\begin{ruledtabular}
\begin{tabular}{ccc}
$E$ & degeneracy& decomposition  \\
\hline
0 & 1 & \\
0.321792450 & 1& $\fb_1$ \\
 0.643584900 & 1& $2\fb_1$  \\
 0.942155004 &2& $\ff_1$ \\
0.965377350 &1& $3\fb_1$ \\
 1.263947454 &2& $\ff_1 + \fb_1$ \\
1.287169800 &1& $4\fb_1$ \\
1.585739904 &2& $\ff_1 + 2\fb_1$ \\
1.608962251 &1& $5\fb_1$ \\
1.884310009 &4& $2\ff_1$ \\
1.907532354 &2& $\ff_1 + 3\fb_1$ \\
1.930754703 &1& $6\fb_1$
\end{tabular}
\end{ruledtabular}
\label{tableone}
\end{table}

As an example, Fig.\ \ref{levelflowk} shows the NRG flow of the low-lying eigenenergies of ${H}_N$ as a function of even iteration number $N$, for fixed Kondo coupling $J$ and five different couplings $g$ to a sub-Ohmic bosonic bath with $s=0.2$.  Increasing $N$ corresponds to considering the problem on a sequence of exponentially decreasing energy scales [Eq.\ (\ref{tempdef})]. The bosonic chain was extended at each even iteration starting at $N=4$. Figure \ref{levelflowk}(a) shows the flow for $g=0$, where the impurity spin and bosonic bath are decoupled.  The fixed-point spectrum reached, with effective $J$ infinite, corresponds to that of the regular Kondo model \cite{Wilson:75}  plus a chain of free bosons.  The resulting fixed-point Hamiltonian, ${H}_K^*=\hat{R}^2[H_K^*]$,  thus consists of a chain of free bosons 
 and a  chain of free fermions described by
\be
H^{(L)}_{F,N}=\alpha^{-1}\Lambda^{N/2}\sum_{n=L}^{N}\epsilon_n f_{n\sigma}^{\dagger}f_{n,\sigma}+\tau_n(f_{n\sigma}^{\dagger}f_{n-1,\sigma}+ \mbox{H.c.}).
\label{freefermi}
\ee
with $L=1$, reflecting the fact that the local moment and conduction electron state at the impurity ($f_0$) are so strongly coupled --- via an infinite Kondo interaction $J$ --- that those degrees of freedom are frozen out.  Each chain problem can be numerically diagonalized to yield a set of single-particle levels that does not depend on $N$ for $N\gg 1$ (apart from an odd-even alternation for the fermionic case).  A decomposition of the lowest lying eigenstates of the even-$N$ Kondo fixed point is provided in Table \ref{tableone}.  

\begin{figure}
\begin{center}
\epsfig{file=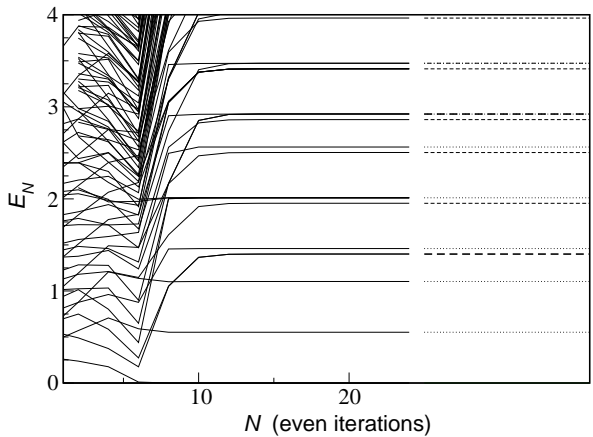, width=7cm, angle=270}
\end{center}
\caption{Interpretation of the low-lying energy levels of the localized fixed-point Hamiltonian $H_L^*$.  The left-hand portion of the figure shows $H_N$ as a function of even iteration number $N$, for $s=0.2$, $J=0.5$, and $g=0.7$.  The spectrum can be decomposed into the product of  pure-fermionic and pure-bosonic spectra.  Plotted on the right-hand side are the levels listed in Table \protect \ref{tableone}, with pure-fermionic excitations shown as dotted lines, pure-bosonic excitations as thick dashed lines, and mixed excitations shown as  thin dashed lines.} 
\label{lmlevels}
\end{figure}

For all $g<g_c$, the effective bosonic coupling flows to zero [see Fig.\ \ref{flow}(a)], while the effective $J$ flows to infinity, and the fixed-point level structure reached for large iteration number $N$ is identical to that for $g=0$.  This is seen clearly in Fig.\ \ref{levelflowk}(b).  However, for $g$ close to $g_c$, the flow first approaches an unstable or critical fixed point. This is also clearly observed in Fig.\ \ref{levelflowk}(b) for $10<N<20$, before the flow crosses over to the stable Kondo fixed point.  The low-energy scale set by the crossover from unstable to stable fixed point vanishes as $g\ra g_c$ and is discussed further in Sec.\ \ref{seccross}.  The level structure for the critical fixed point is shown in Fig.\ \ref{levelflowk}(c), obtained for $g\simeq g_c$.

\begin{table}[b]
\caption{\label{tab:table2}  Low-lying eigenstates of the even-$N$ fixed-point Hamiltonian $H_L^*$, which can be decomoposed into the product of a pure bosonic and pure fermionic spectrum.  The former, $\{\sb_i\}$, is that of the localized fixed point for the spin-boson model and is independent of $g>g_c$.   The latter  $\{\ffv_i\}$ correspond to a pure-fermionic anisotropic Kondo model:  $J^*_z(g=0.7)\simeq 2.772$ and $J^*_{\perp}=0$.  The fixed-point value $J^*_z$ diverges as $g\ra g_c^+$.}
\begin{ruledtabular}
\begin{tabular}{ccc}
$E$ & degeneracy& decomposition  \\
\hline
    0.000000        	&2&\\ 
    0.551499 	&2& $\ffv_1$	\\
     1.102998 	&2& $2\ffv_1$	\\
    1.400966  	&2& $\sb_1$	\\
     1.460443 	&2& $\ffv_2$	\\ 
    1.952465	&2&  $\sb_1+\ffv_1$	\\
   2.011942  	&4&  $\ffv_1+\ffv_2$	\\
   2.503964 		&2&	$\sb_1+2\ffv_1$	\\
   2.563441 		&2&	$2\ffv_1+\ffv_2$\\
   2.861409 		&2&	$\sb_1+\ffv_2$\\
    2.920886 	&2&	$2\ffv_2$\\
   2.922716 		&2&	$\sb_2$\\
    3.412908  	&4&	$\sb_1+\ffv_1+\ffv_2$\\
    3.472385 	&2&	$\ffv_1+2\ffv_2$\\	 
    3.474215 	&2&	$\sb_2+\ffv_1$\\ 
   3.964407  	&2&	$\sb_1+2\ffv_1+\ffv_2$\\ 
\end{tabular}
\end{ruledtabular}
\label{tabletwo}
\end{table}

In the localized phase $g>g_c$, the spectrum of the fixed-point Hamiltonian $H_L^*$ can also be decomposed into a product of pure-bosonic and pure-fermionic spectra.  
Figure \ref{lmlevels} shows NRG level flows for $g=0.7$ and otherwise the same parameters as Fig.\ \ref{levelflowk}.  Localized-fixed-point behavior is there seen for $N\gtrsim 15$, and the fixed-point spectrum can be decomposed into pure-fermionic excitations, pure-bosonic excitations, and appropriate linear combinations of the two, as shown in Table \ref{tabletwo}.
\begin{figure}
\begin{center}
\epsfig{file=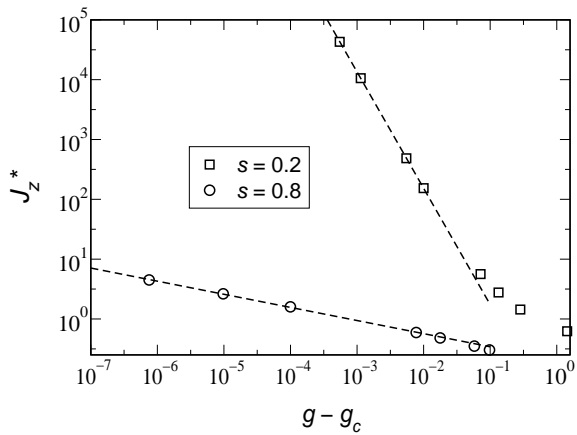, width=6.75cm, angle=270}
\end{center}
\caption{Effective coupling $J_z^*$ at the localized fixed point for the bath exponents specified in the legend and $J=0.5$.  $J_z^*$ exhibits a power-law divergence as $g\ra g_c^+$ [see Eq.\ (\protect\ref{effjz}]. The values of the exponent $\beta$ agree with those obtained in Sec.\ \protect\ref{secresponse} to within 2\%.} 
\label{locjzeff}
\end{figure}
The bosonic spectrum is identical to that of the localized fixed point of the corresponding sub-Ohmic spin-boson model, while the fermionic spectrum corresponds to that of an anisotropic Kondo model in which $J_{\perp}$ has renormalized to zero [see Fig.\ \ref{flow}(a)].  The fixed-point value of $J_z$ is finite for $g>g_c$ and diverges as $g\ra g_c^+$ according to
\be
J^*_z\propto (g-g_c)^{-\beta},
\label{effjz}
\ee
where $\beta$ is the magnetic critical exponent defined in Sec.\ \ref{secresponse}.  This behavior is demonstrated in Fig.\ \ref{locjzeff}, which shows $J^*_z$ versus $g-g_c$ for $J=0.5$ and two values of the bosonic bath exponent.  In both cases the values of the exponent $\beta$ agree with those obtained in Sec.\ \ref{secresponse} to within 2\%.

As $s\ra 0^+$ the critical fixed point merges with the Kondo fixed point.  For any $-1<s\le 0$, the critical coupling $g_c(s)=0$, i.e., the system is localized for any nonzero $g$. For small $g$ the flow approaches the Kondo fixed point discussed above [shown schematically in Fig.\ \ref{flow}(b)], before crossing over to the localized fixed point on an energy scale $T_*$ that vanishes as $g\ra 0$.  For $s=1$ the QPT is Kosterlitz-Thouless-like, considered in Sec.\ \ref{s1section}, while for $s>1$ the Kondo fixed point is reached regardless of the strength of coupling to the bath.

\subsection{Critical coupling}
\label{seccrit}
\begin{figure}
\begin{center}
\epsfig{file=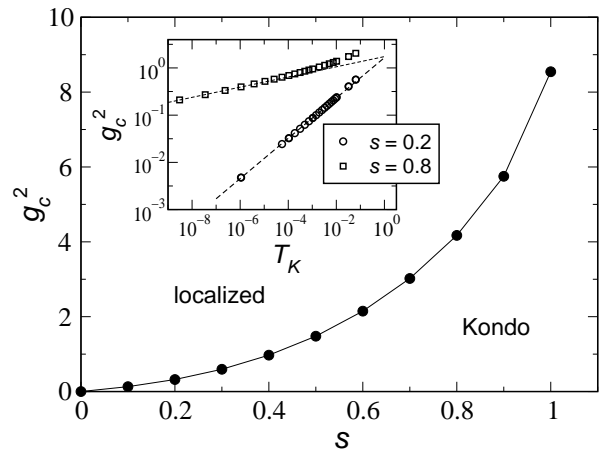, width=6.5cm, angle=270}
\end{center}
\caption{Critical coupling $g_c^2$, separating Kondo and localized phases, versus bosonic bath exponent $s$ for fixed Kondo interaction $J=0.5$.   The inset shows the $\tkn$ dependence of $g_c^2$ for representative bath exponents $s=0.2$ and $0.8$ and the dashed lines are fits to the form $g_c^2\propto \tkn^{1-s}$, which is followed for $\tkn \ll 1$.  See text for discussion.  }
\label{boundary}
\end{figure}

The critical coupling $g_c$, which marks the phase boundary between Kondo-screened and localized phases for $0\le s \leq 1$, is shown in Fig.\ \ref{boundary} as a function of bath exponent $s$.
  For $s\ra 0^+$, $g_c^2(s)$ vanishes linearly in $s$.   The line of continuous quantum phase transitions for $0<s<1$ terminates at the Kosterlitz-Thouless transition point at $s=1$. 

\begin{figure}
\begin{center}
\epsfig{file=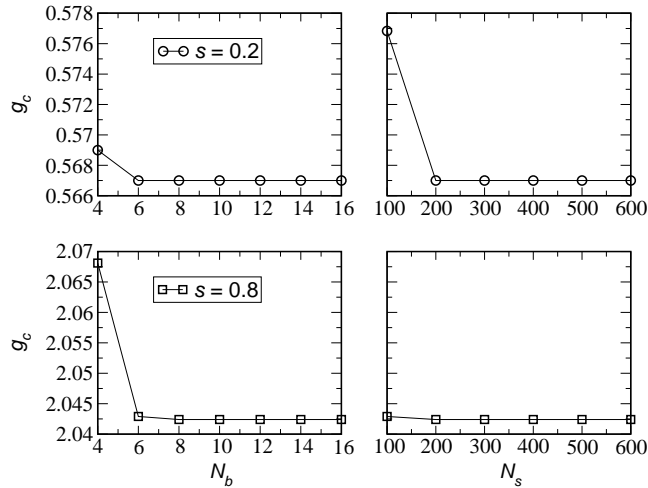, width=6.75cm, angle=270}
\end{center}
\caption{Dependence of critical coupling $g_c$ on NRG truncation parameters $N_b$ and $N_s$ for $J=0.5$,  NRG discretization $\Lambda=9$, and bath exponents  $s=0.2$ and $s=0.8$.  The first column demonstrates rapid convergence in $N_b$ for fixed $N_s=500$.  The second column shows convergence in $N_s$ for fixed $N_b=8$.}
\label{gcofnbns}
\end{figure}
Figure \ref{gcofnbns} demonstrates, for representative bath exponents $s=0.2$ and $0.8$, that the critical couplings converge rapidly upon increasing the truncation parameters $N_s$ and $N_b$, denoting respectively the number of states retained from one NRG iteration to the next and the maximum number of bosons allowed per site of the bosonic chain.  In percentage terms, $g_c$ converges with increasing $N_b$ more rapidly for smaller $s$, but converges with increasing $N_s$ more rapidly for larger $s$.  However, in all cases the choices $N_s=500$ and $8<N_b<12$ are sufficient  for the determination of phase boundaries for $\Lambda=9$.  
In Sec.\ \ref{secresponse} we show that critical exponents are also adequately converged for these parameters, a result of crucial importance in judging the  success of the approach, and one  that is not \textit{a priori} guaranteed.

Of course, deep inside the localized phase it is to be expected that an approach using the lowest $N_b+1$ eigenstates of $b_M^{\dagger}b_M$ as basis states for each bosonic site $M$, will ultimately fail due to a divergence of the mean site-occupancy.  In a recent study of the SBM using a bosonic NRG approach,\cite{Bulla:03,Bulla:04} Bulla \textit{et al}.\ have employed a bosonic basis of displaced oscillator states implemented within an alternative ``star'' formulation of the NRG.  While this approach describes well the localized fixed point, it fails to capture correctly both the flow to the delocalized fixed point and the critical properties of primary interest.\cite{Bulla:04}  

\begin{figure}
\begin{center}
\epsfig{file=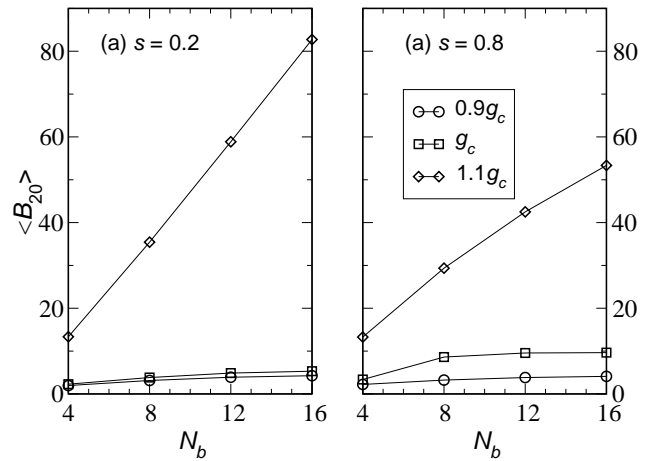, width=6.75cm, angle=270}
\end{center}
\caption{Dependence of $\bra\hat{B}_{20}\ket$, defined in text and evaluated at the characteristic temperature scale of iteration $N=20$, on bosonic truncation parameter $N_b$ for (a) $s=0.2$ and (b) $s=0.8$. Data were obtained for $J=0.5$ and the bosonic couplings $g$ shown in the legend.
Convergence with respect to $N_b$ is reached rapidly for $g\le g_c$.}
\label{aa}
\end{figure}

The preceding comments are well-illustrated by Fig.\ \ref{aa}, which considers the operator for the total number of bosons at iteration $N$,
\be
\hat{B}_N=\sum_i^{M(N)} b^{\dagger}_ib_i,
\ee
and its expectation value evaluated at the characteristic temperature $T_N$, Eq.\ (\ref{tempdef}).  Here, $M(N)$ denotes the highest labelled bosonic site at iteration $N$.  Figure \ref{aa} shows $\bra \hat{B}_{20}\ket$ versus bosonic truncation parameter $N_b$ for both Kondo ($g<g_c$) and localized ($g>g_c$) phases and at the QCP ($g=g_c$) itself.  In the Kondo phase and at the QCP,  expectation values converge rapidly in $N_b$, indicating that the lowest eigenstates of $H_{N}$ are well-described using only the lowest $N_b+1$ eigenstates of $b^{\dagger}_M b_M$ as basis states for each bosonic site $M$.  It is clear that no such convergence is achieved inside the localized phase.   The difference between the saturated values of $\bra \hat{B}_{20}\ket$ for critical and delocalized phases is smaller for lower $s$,  consistent with the merging of the critical and delocalized fixed points as $s\ra 0$ (see Fig.\ \ref{flow}).

The aim of the present work is to provide a thorough numerical account of the critical behavior of the BFKM.  As is demonstrated explicitly in  Sec.\ \ref{secresponse}, the approach described in Sec.\ \ref{secnrg} provides such a description.   We emphasize that critical properties are well-captured regardless of the phase --- Kondo \textit{or} localized --- from which the critical point is approached.

We now return to a discussion of the phase boundary between Kondo and localized regimes of the model.  Reducing $J$, and hence the Kondo scale $\tkn$ of the pure-fermionic ($g=0$) problem, squeezes out the Kondo-screened phase of the BFKM; specifically, we find
\be
g_c^2(s)=c(s)~\tkn^{1-s} \mbox{\ for \ } \tkn\ll 1,
\label{goftk}
\ee
with $c(s)$ an $s$-dependent constant that vanishes both as $s\ra 0^+$ and $s\ra
 1^-$, in agreement with perturbative (in $\epsilon\equiv 1-s$) RG calculations. 
\cite{Zhu:02} 

The form of Eq.\  (\ref{goftk}) is demonstrated in the inset to Fig.\ \ref{boundary}, which shows $g_c^2$ versus $\tkn$ on a logarithmic scale for $s=0.2$ and $0.8$.  
Here, we define $\tkn\equiv D\exp[-A(\Lambda)/J]$. The $\Lambda$-dependent constant $A(\Lambda)$ tends to correct the effects of discretization, which reduce the effective coupling between impurity and band, and leads to the most faithful description of the impurity problem.\cite{Oliveira:81}
The slopes of the linear fits, shown as dashed lines, reveal that the exponent is $1-s$ to high accuracy, which relation holds for all $0<s<1$.  We note that this exponent follows from simple dimensional arguments.  When all other energy scales in the problem are much smaller than the cut-offs, the problem may be rescaled  in terms of $\tkn$, with a single dimensionless parameter $g^2(\tkn/\w_0)^{1-s}$.

If the Kondo scale $\tkn$ is recognized as a (renormalized) tunneling amplitude between the two spin states of the impurity, then Eq.\ (\ref{goftk}) follows from the sub-Ohmic spin-boson model result,\cite{Bulla:03} where it is found that the critical dissipation strength $\alpha_c\propto \Delta^{1-s}$ for $\Delta\ll \w_0$.  Such an interpretation follows, of course, from the bosonization discussed in Sec.\ \ref{mapping}.  The sub-Ohmic BFKM can be mapped to a spin-boson model with a sub-Ohmic bath \textit{plus} an Ohmic contribution arising from bosonization of the Kondo part of the Hamiltonian.  While the sub-Ohmic behavior dominates the bath spectral function at low energies, and hence determines the universal critical properties, the Ohmic contribution acts to renormalize the bare couplings, as discussed more generally in Ref.\ \onlinecite{Leggett:87}.  Specifically $J\ra \tkn$, leading to an effective SBM with $\Delta\propto \tkn$ and a sub-Ohmic bath.

\subsection{Crossover scale}
\label{seccross}
We now extract the correlation length exponent $\nu$ for the BFKM, considering bath exponents in the range $0<s<1$.  It is a general feature of a  continuous quantum phase transition \cite{Sachdev:99} that a characteristic energy scale for fluctuations above the ground state vanishes as $g\ra g_c$ with a universal exponent $\nu z$.  The exponent $\nu$ describes the divergence of the spatial correlation length at the critical coupling, and the dynamical exponent $z$ specifies the effective spatial dimensionality of each time dimension.  Note that $z=1$ for impurity models.

\begin{figure}
\begin{center}
\epsfig{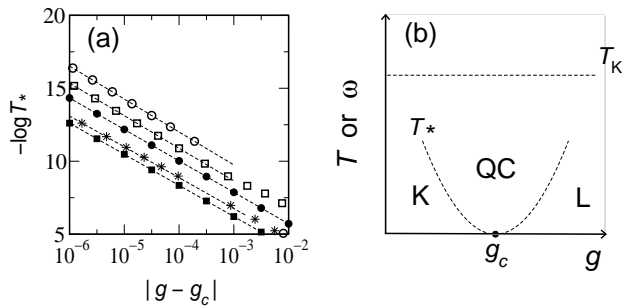}
\end{center}
\caption{(a) Power-law behavior of the crossover scale $T_*\propto\Lambda^{-N_*/2}$ in the vicinity of the QCP.  Data are shown for $s=0.5$, $J=0.5$ (solid symbols) and $J=0.1$ (open symbols), for both the Kondo (circles) and localized (squares) phases with NRG discretization $\Lambda=9$. Kondo-phase data obtained for $s=0.5$, $J=0.5$, and $\Lambda=3$ are shown as stars.  The correlation-length exponent $\nu(s)$ is universal and is independent of $J$, of the phase from which the QCP is approached, and of $\Lambda$.  (b) Schematic crossover diagram valid near the QCP for $0<s<1$.  The dashed line shows $T_*$, the scale for crossover between critical and stable fixed-point behavior in static or dynamical quantities. The horizontal dashed line marks $\tkn$, the Kondo scale of the pure-fermionic problem, which serves as a high-energy cutoff for the quantum-critical behavior.}
\label{crossover}
\end{figure}

\begin{figure}
\begin{center}
\epsfig{file=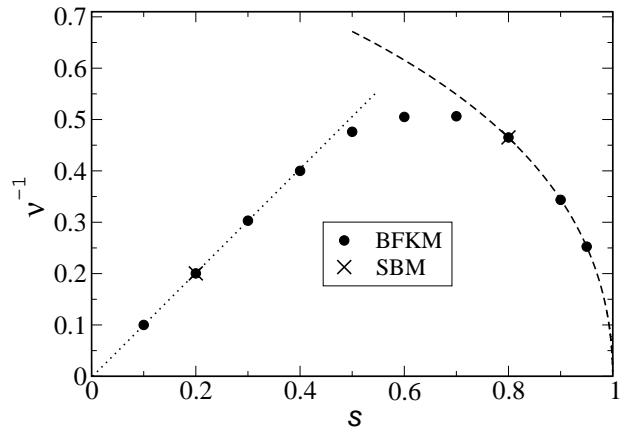, width=7cm, angle=270}
\end{center}
\caption{Dependence of the correlation length exponent $\nu(s)$ on bath exponent $s$.  Results for the spin boson model are plotted as crosses.  For $s\ra 0$, $\nu^{-1}\simeq s$ (fit shown as dotted line) while $\nu^{-1}$ vanishes as $\sqrt{2(1-s)}$ for $s\ra 1$ (fit shown as dashed line).}
\label{expon_nu}
\end{figure}
Close to the critical coupling $g_c$, the crossover  from the unstable quantum-critical regime to one or other of the  stable regimes, defines an energy scale $T_*$ that vanishes precisely at $g_c$ in a fashion governed by the correlation-length  exponent:
\be
T_*\propto |g-g_c|^{\nu}.
\label{nudef}
\ee
 $T_*^{-1}$ sets the time-scale for the decay of the local-moment fluctuations.  Close to the QCP the characteristic energy scale in the problem is $T_*$, while at the QCP itself the only relevant energy scale is the temperature, as discussed in Sec.\ \ref{sectionds} in the context of $\w/T$ scaling.  The crossover scale $T_*$ may be determined directly from the NRG many-body spectrum, via $T_*\propto \Lambda^{-N_*/2}$.  Here, $N_*$ is the NRG iteration number at which a selected energy level crosses over from the critical fixed point to a stable fixed point. For example, for the lowest nonzero energy level in Fig.\ \ref{levelflowk}(b), the clear crossover from critical ($E_{c}\simeq 0.246$) to Kondo regime ($E_{K}\simeq 0.322$) occurs for $N_*\simeq 27.2$, with $N_*$ here defined such that ${E_N}_*=\frac{1}{2}(E_{c}+E_{K})$.  Alternatively, $T_*$ may be extracted from crossovers  seen directly in properties such as the local static or dynamical susceptibility or the single-particle spectrum.  Good agreement is found in all cases to within the numerical uncertainty, which is typically less than 1\%.

Fig.\ \ref{crossover}(a) shows $T_* \propto N_*$ versus $|g-g_c|$  for $J=0.5$, $s=0.5$ and a range of couplings $g$ approaching $g_c\simeq 1.22$ from either phase.  The form Eq.\ (\ref{nudef}) holds well over several orders of magnitude, as reflected in the linear behavior of the data on a logarithmic scale, and any deviations for small $|g_c-g|$ are removed by a more accurate determination of $g_c$.  The correlation-length exponent $\nu$ is proportional to the slope of the lines and it is therefore clear to see  that $\nu$ is independent of Kondo coupling $J$ and the phase (Kondo or localized) from which the QCP is approached. The exponent is universal, as are all others discussed below, and depends only on the bosonic bath exponent $s$, which thereby plays the role of a dimension.
The insensitivity to $\Lambda$ is also demonstrated in Fig.\ \ref{crossover}(a), justifying our choice to work with $\Lambda=9$ (and hence fewer retained states) for extracting critical exponents.

The behavior shown in Fig.\ \ref{crossover}(a) is representative of that for all $s$ in the range $0< s < 1$. 
A generic crossover diagram is shown in Fig.\ \ref{crossover}(b), valid in the vicinity of the QCP and for $T$ (and $\w$) below a nonuniversal cutoff scale (above which free-local-moment behavior obtains).   For $T<T_*$, behavior characteristic of either the Kondo ($g<g_c$) or the localized ($g>g_c$) phase is observed, while for $T>T_*$ behavior characteristic of the QCP itself obtains.  At the QCP, where $T_*=0$, the latter behavior persists right down to $T=0$.  This results in the familiar picture for a quantum phase transition of quantum-critical behavior emanating from a singular point at $T=0$ and influencing a large part of the phase diagram in the $g$-$T$ (or $g$-$\w$) plane.  Such a picture has emerged in recent years in a number of quantum impurity problems. \cite{Vojta:04a}

The dependence of $\nu$  on the continuous parameter (dimension) $s$ is shown in Fig.\ \ref{expon_nu}(a).   For both $s\ra0^+$ and $s\ra 1^-$ the correlation length exponent diverges, which reflects the qualitative changes occuring at these critical dimensions, as described above.  For small $s$,
\be 
\nu(s)\simeq 1/s, 
\label{nulows}
\ee
 which result holds reasonably well for $s=0.4$, and becomes asymptotically exact as $s\ra 0^+$ [see dotted line fit in Fig.\ \ref{expon_nu}(a)].  For $s\ra 1$ the exponent approaches
\be
\nu(s)\simeq 1/\sqrt{2(1-s)}+C,
\label{nusto1}
\ee
with $C\simeq 0.4$, which fit is shown as the dashed line in Fig.\ \ref{expon_nu}(a).   Most importantly, we demonstrate that the correlation length exponents of this Ising anisotropic BFKM are in essentially exact agreement with those of the spin-boson model obtained in Refs.\ \onlinecite{Bulla:03} and \onlinecite{Vojta:04}.  This is true for all exponents calculated, providing numerical confirmation that the quantum-critical points of the two models belong to the same universality class --- an equivalence implied by the bosonization treatment discussed in Sec.\ \ref{mapping}.

\begin{figure}
\begin{center}
\epsfig{file=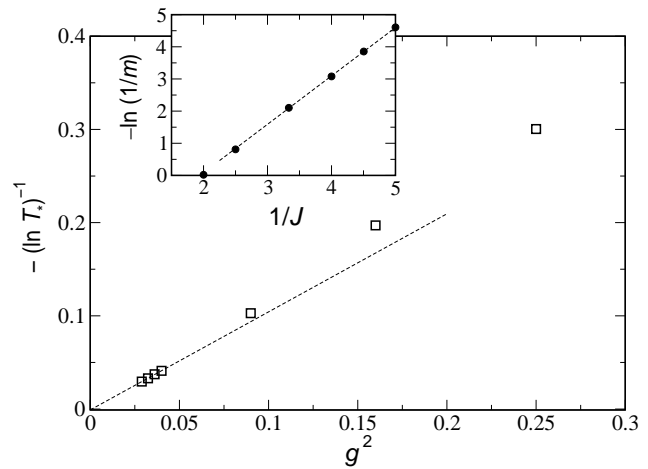, width=6.75cm, angle=270}
\end{center}
\caption{Crossover scale for $s=0$ vanishing  as Eq.\ (\protect\ref{s0scale}) for $g\ra 0$. The data in the main figure were obtained for $J=0.5$.  A linear fit to the  data points at lowest $g^2$ is shown as a dashed line.  The inset demonstrates that the slope $m$ of the fit varies as $1/\tkn$ --- see text.}
\label{s0}
\end{figure}

The results $\nu(s)\simeq 1/s$ and $g_c^2/T_{\K}\propto s$ as $s\ra 0$ [Eqs.\ (\ref{nulows}) and (\ref{goftk}) respectively] imply a form for the vanishing of the crossover scale $T_*$ at $s=0$.  Here the system is always localized for any $g>0$ and we predict that $T_*$, describing the direct crossover from Kondo to localized behavior, vanishes approaching $g=g_c=0$ as
\be
\ln T_*\propto  -\frac{T_{\K}}{g^2}, \ \ \ \ \mbox{for } s=0.
\label{s0scale}
\ee
This is confirmed numerically in Fig.\ \ref{s0}, which shows data for $J=0.5$ rescaled according to Eq.\ (\ref{s0scale}).  The dashed line shows a linear fit for $g^2 \ll 1$.  The inset demonstrates that the slope of the fit is proportional to $\tkn^{-1}$, where $\tkn\equiv D\exp[-A(\Lambda)/J]$.   For the sub-Ohmic spin-boson model, identifying $\Delta\propto \tkn$ and $\alpha\propto g^2$ leads to $\ln T_*\propto (-\Delta/\alpha)$ for $s=0$, which result has also been obtained via a perturbative RG analysis. \cite{Vojta:04}

\subsection{Local magnetic response}
\label{secresponse}
The local susceptibility, $\cl(\w)=\cl'(\w)+i\cl''(\w)$, is defined as
\be
\cl(\w)=i\int_0^{\infty}e^{-i\w t}\bra[S_z(t), S_z(0)]\ket
\ee
with $\cl'(\w=0)$ being the static response
\be
\cl(0) = -\left.\frac{\partial\bra S_z\ket}{\partial h}\right|_{h=0}=\lim_{h\ra 0} -\frac{\bra S_z\ket}{h}
\label{clzero}
\ee
 to a local magnetic field $h$ acting only at the impurity site  through an additional term in the Hamiltonian $\Delta \hat{H}= h S_z$ (setting $g\mu_B\equiv 1$).

In the present section we focus on the sub-Ohmic case with $0<s<1$.  The case of Ohmic dissipation, $s=1$, is discussed separately in Sec.\ \ref{s1section}.
In Sec.\ \ref{statics} we consider the local static susceptibility $\cl(T; \w=0)$ calculated by evaluating the right-hand side of Eq.\ (\ref{clzero}) directly for small fields in the range $10^{-12}\le h \le 10^{-8}$.  We extract various critical exponents for the transition which are shown to satisfy hyperscaling relations characteristic of an interacting quantum-critical point. 
We calculate the dynamical local susceptibility in Sec.\ \ref{clomega}  which obeys $\w/T$ scaling in the vicinity of the QCP.  We reiterate that all results presented are converged with respect to the truncation parameters $N_b$ and $N_s$.  For the data discussed, $N_b=8$ and $N_s=500$ are sufficient to converge critical exponents to the required accuracy --- see, e.g., Fig.\ \ref{expons}.

\subsubsection{Static susceptibility}
 \label{statics}
\begin{figure}
\begin{center}
\epsfig{file=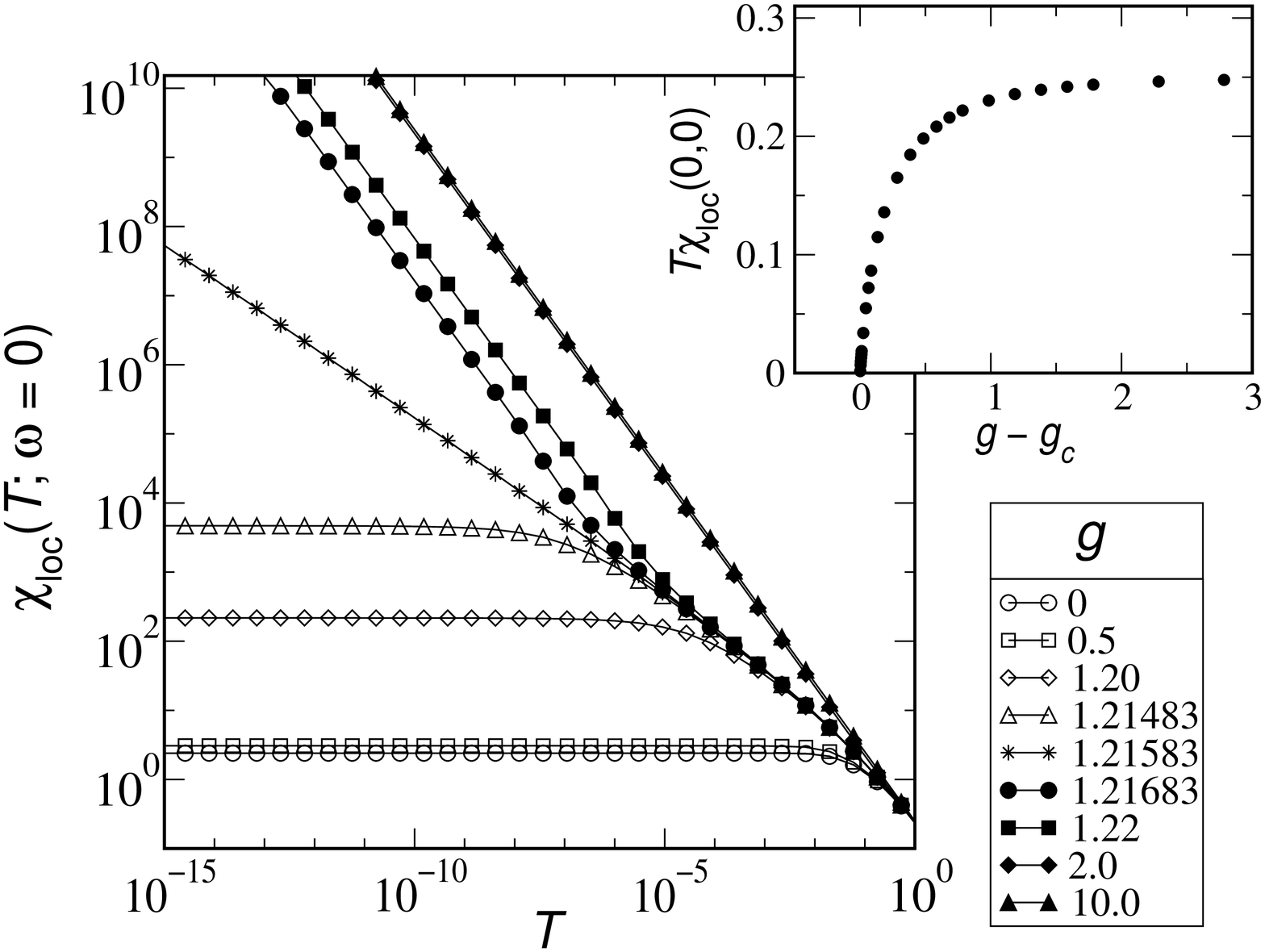, width=6.75cm, angle=270}
\end{center}
\caption{Static local susceptibility $\cl(T;\w=0)$ versus $T$ for $s=0.5$, $\rho_0J=0.5$ and different $g$, where $g_c\simeq 1.21583$.  $\cl(0,0)=0$ for $g<g_{c}$. At the critical coupling we find $\cl(T;0)\propto T^{-x}$ with $x=0.500(2)$.  The inset shows the vanishing of $T\cl(0,0)$ for $g\ra g_c^+$.    Analogous data for $s=0.2$ are shown in Fig.\ 2 of Ref.\ \protect \onlinecite{Glossop:05a}.}
\label{chilocvst}
\end{figure}

The Kondo and localized phases are readily distinguished by the limiting low-temperature behavior of $\cl(T; \w=0)$.
If a local moment $M_{\mbox{\ssz{loc}}}\equiv\bra S_z\ket$ exists at the impurity site for $T\ra 0$, then
\be
\cl(T; \w=0)=\frac{M_{\mbox{\ssz{loc}}}(0)^2}{T} \mbox{\ for\ } T\ll T_*.
\label{curie}
\ee
We find such behavior for $g>g_{\cc}$, as shown in Fig.\ \ref{chilocvst}.

For $g\gg g_{\cc}$ the residual moment $M_{\mbox{\ssz{loc}}}(0)$ approaches the value for a completely decoupled spin,  $M_{\mbox{\ssz{loc}}}(0)^2=S(S+1)/3$, as seen clearly in the inset to Fig.\ \ref{chilocvst}.  For $0<s<1$, the low-temperature limit of $T\cl(g>g_{\cc}, T; \w=0)$   vanishes continuously at the QCP, viz 
\be
\mbox{lim}_{T\ra 0}T\cl(g>g_{\cc}, T; \w=0)\propto (g-g_{\cc})^{\lambda},
\label{lamdef}
\ee
as is also clear from the inset to Fig.\ \ref{chilocvst}. Equations (\ref{curie}) and (\ref{lamdef}) imply that 
\be
M_{\mbox{\ssz{loc}}}(g>g_c;T=0)\propto (g-g_c)^{\beta} 
\label{betadef}
\ee
with $\beta=\lambda/2$.
\begin{figure}
\begin{center}
\epsfig{file=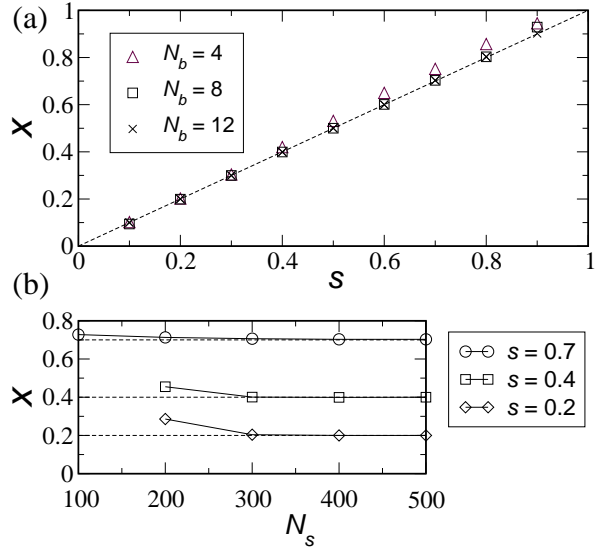, width=7.5cm, angle=270}
\end{center}
\caption{Dependence of the critical exponent $x$, as defined in Eq.\ (\protect\ref{xdef}), on  bath exponent $s$ in the range $0<s<1$. The data were obtained for Kondo coupling $J=0.5$, discretization parameter $\Lambda=9$, and (a) $N_s=500$, (b) $N_b=12$.  The dotted lines show $x=s$.}
\label{expons}
\end{figure}

\begin{figure}
\begin{center}
\epsfig{file=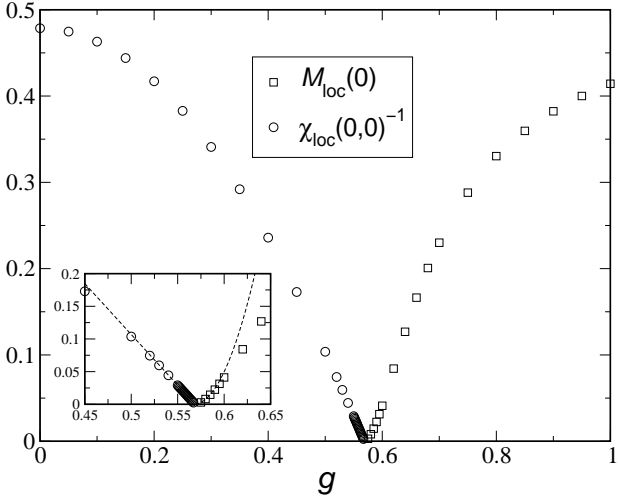, width=7cm, angle=270}
\end{center}
\caption{Order parameter $M_{\mbox{\ssz{loc}}}(g>g_c; T=0)$ and renormalized Kondo temperature $\cl(g<g_c; T=0,\w=0)^{-1}$ versus bosonic coupling $g$, for $s=0.2$ and $J=0.5$.  Both quantities vanish continuously for $g \ra g_c$. The inset shows fits (dashed lines) for small $|g-g_c|$.  See text for discussion.}
\label{s0.2gammabeta}
\end{figure}

For $g<g_{\cc}$, by contrast, we find the characteristic signature of a screened ground state with quenched impurity degrees of freedom (as exists ubiquitously in the regular Kondo model):
\be
\cl(T; \w=0)= \mbox{const.} \ \ \ \mbox{for\ }  T\ll T_*,
\ee
which behavior is also seen clearly in Fig.\ \ref{chilocvst}.
For $0<s<1$ it is found that 
\be
\cl(g<g_c; T=0, \w=0)\propto (g_{\cc}-g)^{-\gamma},
\label{gammadef}
\ee
signalling a collapse of the Kondo effect at $g=g_c$ (see Fig. \ref{s0.2gammabeta}) discussed further below.

At the QCP itself, $g=g_{\cc}$, we find the anomalous behavior
\be
\cl(T;\w=0, g=g_c)\propto T^{-x}.
\label{xdef}
\ee
To high accuracy we find
\be
\label{xeqs}
x=s 
\ee
for $0<s<1$, confirming the prediction of $\epsilon$-expansion studies \cite{Zhu:02,Zarand:02} that $x=s$ to all orders in $\epsilon\equiv 1-s$.  The result Eq.\ (\ref{xeqs}) has also recently been reported for the sub-Ohmic spin-boson model. \cite{Bulla:03,Vojta:04}   We note in passing that since the total spin quantum number is not conserved in the BFKM, the impurity contribution to the total susceptibility $\chi_{\mbox{\scriptsize{imp}}}$ also acquires an anomalous
exponent at criticality.  Again, we find  $\chi_{\mbox{\scriptsize{imp}}}(T)\propto T^{-s}$.

For $T\ll \tkn$ the local static susceptibility obeys the scaling form
\be
\tkn\cl(T;\w=0)=\left(\frac{T_*}{\tkn}\right)^{-s}\bar{\phi}_{s,p}\left(\frac{T}{T_*}\right)
\label{staticscaling}
\ee
where $\bar{\phi}_{s,p}$ is a universal function and $p=K$ or $L$ denotes the phase.  For $z\gg 1$, $\bar{\phi}_{s,p}(z)\propto z^{-s}$ and  $\tkn\cl(T)\propto (T/\tkn)^{-s}$ is independent of $T_*$.

Before proceding we illustrate the convergence of the critical exponents in the NRG truncation parameters $N_b$ and $N_s$, which denote respectively the maximum number of bosons allowed per site of the bosonic chain and the maximum number of states retained from one NRG iteration to the next.  The behavior of the critical exponent $x$ is typical and is shown in Fig.\ \ref{expons} as a function of bosonic bath exponent $s$.
 Figure \ref{expons}(a) shows data obtained for $N_b$ = 4, 8 and 12, each converged with respect to  $N_s$, the number of states retained.  Upon increasing $N_b$, the exponent $x$ is found to approach $x=s$ [dashed line in Fig. \ref{expons}(a)] for all $s$, the convergence being faster for small $s$.  For this and all other critical exponents calculated we find that by $8<N_b<12$ sufficient convergence is achieved.  Figure \ref{expons}(b) illustrates the dependence of  $x$ on the number of states retained $N_s$, calculated for $N_b=12$ and three values of $s$. The exponent converges rapidly with increasing $N_s$, approaching $x=s$ (dashed lines) in each case.  In summary, we find that by $N_b=12$ and $N_s=500$ the critical exponent $x$ is converged to within 1\% of $x=s$ for all $s$ examined in the range $0<s<1$.  The observation  that fewer bosons are required to achieve convergence for $s\ll 1$ is consistent with the merging of the critical and delocalized fixed points as $s\ra 0$.

We now return to a discussion of the various critical exponents introduced above, for which similar convergence is observed.
The vanishing  as $g\ra g_c^+$ of the residual moment $M_{\mbox{\ssz{loc}}}(T=0)$,  which serves as the order parameter for the problem, is shown in Fig.\ \ref{s0.2gammabeta} (squares).  This picture remains qualitatively valid for all bath exponents in the range $0<s<1$.  For the specific case $s=0.2$ shown, the exponent defined in Eq.\ (\ref{betadef}) is $\beta\equiv\frac{1}{2}\lambda\simeq 1.99(2)$, with the estimated numerical uncertainty in the last digit given in parentheses. The fit is shown as a dashed line in the inset to Fig.\ \ref{s0.2gammabeta}.  Also shown in Fig.\ \ref{s0.2gammabeta} is the vanishing of $\cl(g<g_c; T=0, \w=0)$ for $g\ra g_c^-$ described by exponent $\gamma$ [Eq.\ (\ref{gammadef}].  $\cl(0,0)^{-1}$ measures the Kondo temperature, which is strongly renormalized due to the competiting coupling to the dissipative bosonic bath, and its vanishing as $g\ra g_c^-$ signals criticality in the Kondo screening.  For $s=0.2$, $\gamma\simeq 1.00(1)$, and the dotted line in the inset shows a linear fit to the $g\ra g_c^-$ data points.

The dependence of $M_{\mbox{\ssz{loc}}}(h;g=g_c, T=0)$ on the local field $h$ defines the magnetic critical exponent $\delta$:
\be
M_{\mbox{\ssz{loc}}}(h;g=g_c, T=0)\propto |h|^{\frac{1}{\delta}}.
\label{deltadef}
\ee
\begin{figure}
\begin{center}
\epsfig{file=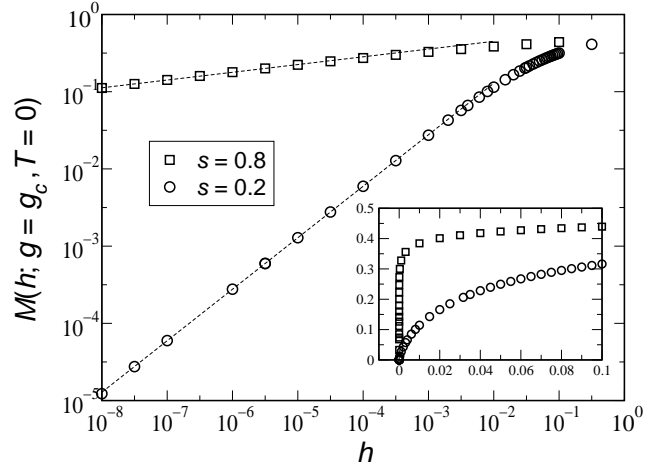, width=6.75cm, angle=270}
\end{center}
\caption{Dependence of residual moment $M_{\mbox{\ssz{loc}}}(g=g_c,T=0)$ on local magnetic field $h$, for $J=0.5$.  The dashed lines are fits to the data, as discussed in the text.  The inset shows the data on an absolute scale.}
\label{deltas0.2}
\end{figure}
Figure \ref{deltas0.2} shows data obtained for $J=0.5$ and $g=g_c$ for both $s=0.2$ and $s=0.8$;  the inset shows the data on an absolute scale, while the main figure reveals the power-law behavior that is observed over several orders of magnitude.  The dashed lines are fits to the $h\le 10^{-3}$ data, of the form of Eq.\ (\ref{deltadef}), with $1/\delta\simeq 0.665(5)$ [0.106(2)] for $s=0.2$ [0.8].

\begin{figure}
\begin{center}
\epsfig{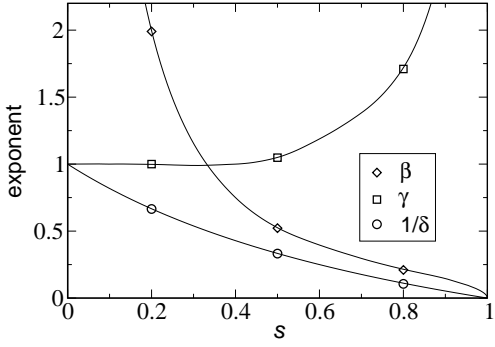}
\end{center}
\caption{Behavior of critical exponents $\beta$, $\gamma$ and $1/\delta$ as a function of bosonic bath exponent $s$ in the range $0<s<1$.  The symbols correspond to directly calculated exponents, while the solid lines were obtained from hyperscaling relations [Eqs.\ (4.23]) using numerical results for $x(s)$ and $\nu(s)$.  Hyperscaling relations are obeyed to within numerical accuracy for all $s$ considered in the range $0<s<1$.}
\label{hyper}
\end{figure}

Hyperscaling relations among the various critical exponents may be deduced starting from the following scaling ansatz for the critical part of the free energy, which assumes that the fixed point is interacting:
\be
F_{\mbox{\ssz{crit}}}=Tf(|g-g_{\cc}|/T^{\frac{1}{\nu}}, |h|/T^b).
\ee
Using the standard thermodynamic relations
\bea
M_{\mbox{\ssz{loc}}}&=&-\frac{\partial F_{\mbox{\ssz{crit}}}}{\partial h},\\
\cl(T; \w=0)&=&-\frac{\partial^2 F_{\mbox{\ssz{crit}}}}{\partial h^2},
\eea
 it readily follows that there are only two independent exponents, i.e.,\
\alpheqn
\bea
\delta&=&\frac{1+x}{1-x}, \label{hyperscal}\\
 \lambda&\equiv&2\beta=\nu(1-x), \\ 
 \gamma &=& \nu x.
\eea
\reseteqn
We find that our numerically  determined exponents for $0<s<1$ obey these relations to high accuracy, as demonstrated in Fig.\ \ref{hyper} which compares numerical obtained exponents $\beta$, $\gamma$ and $1/\delta$ to values predicted by hyperscaling relations.  Numerical values for representative cases are given in Table 1 of Ref.\ \onlinecite{Glossop:05a}.  It follows from the relations Eqs.\ (4.23) and the asymptotic results Eqs.\ (\ref{nulows}) and (\ref{nusto1}) that for $s\ra 0^+$, $1/\delta \simeq 1-2s$, $\lambda\simeq 1/s$ and $\gamma\simeq 1$, while for $s\ra 1^-$, $1/\delta\simeq \frac{1}{2}(1-s)$, $\lambda\propto\sqrt{1-s}$ and $\gamma\propto 1/\sqrt{1-s}$.

\begin{figure}
\begin{center}
\epsfig{file=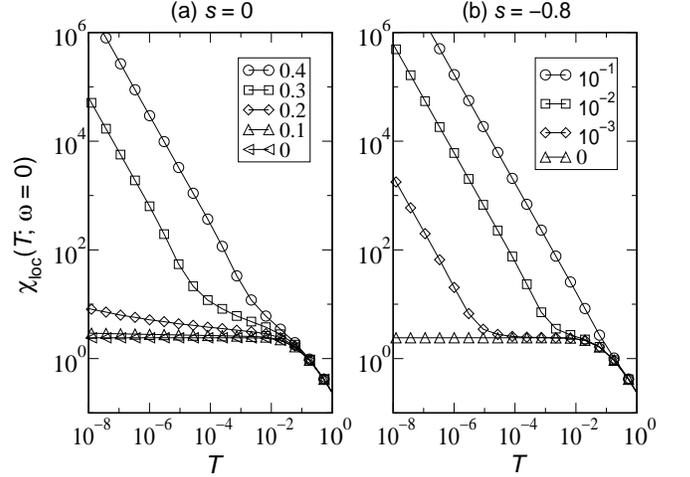, width=6.75cm, angle=270}
\end{center}
\caption{Static local susceptibility $\cl(T;\w=0)$ versus $T$ for $J=0.5$ and (a) $s=0$ and (b) $s=-0.8$.  The legends refer to the bosonic coupling $g$.  Qualitatively similar behavior is observed in either case: as $g\ra 0$ there is a direct crossover from $g=0$ Kondo behavior for $T\gg T_*$, to localized Curie-like behavior for $T\ll T_*$.  The localized-phase crossover scale $T_*$ vanishes as $g\ra 0$ --- see Fig.\ \protect \ref{lambdaslt0}.}
\label{chilocs0}
\end{figure}

\begin{figure}
\begin{center}
\epsfig{file=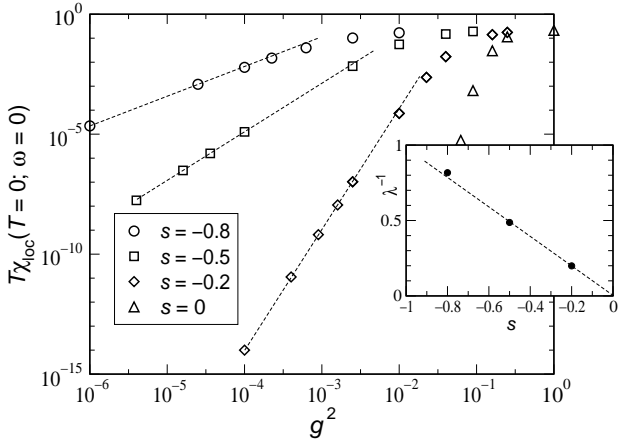, width=6.5cm, angle=270}
\end{center}
\caption{$M_{\mbox{\ssz{loc}}}(T=0)^2=\lim_{T\ra 0}T\cl(T;\w=0)$ versus bosonic coupling $g^2$ for the $s\leq 0$ bath exponents specified in the legend.   For $-1<s\leq 0$ the system is always localized for any $g>0$ and $M_{\mbox{\ssz{loc}}}(T=0)^2$ is found to vanish continuously as $g\ra 0$.
  For $g^2 \ll 1$ the data can be fit to the form $M_{\mbox{\ssz{loc}}}(T=0)^2\propto (g^2)^{\lambda}$.  The inset shows the exponent $\lambda$ versus $s$, which diverges as $\lambda\simeq 1/s$ for $s\ra 0^-$.  See text for discussion.}
\label{lambdaslt0}
\end{figure}

We now turn briefly to the regime $-1<s \le0$ throughout which the system is in the localized phase for any $g>0$.  Figure \ref{chilocs0} shows the static local susceptibility $\cl(T;\w=0)$ versus $T$ for bath exponents $s=0$ and $s=-0.8$.  The observed behavior is typical of any $s$ in the range $-1<0\le 0$.  As indicated by the flow diagram Fig.\ \ref{flow}(b), the $g=0$ (Kondo) spectrum is followed down to progressively lower temperatures as $g\ra 0$ before crossing over directly to localized Curie-like behavior, Eq.\ (\ref{curie}).  Consequently, for $T_*\ll \tkn$, $\cl(T;\w=0)$ obeys the simple scaling form 
\be
\tkn\cl(T;\w=0)=\phi_{s,L}\left(\frac{T}{T_*}\right)
\ee
with no  $T_*$-dependent prefactor, and $\nu=\lambda$.  Figure \ref{lambdaslt0} shows $M_{\mbox{\ssz{loc}}}(0)^2$ as a function of $g^2$ for the $s$ values specified.  In all cases we find that $M_{\mbox{\ssz{loc}}}(0)^2$ vanishes continuously with
\be
M_{\mbox{\ssz{loc}}}(0)^2\propto (g^2)^{\lambda}.
\ee
The inset plots the inverse of the exponent $\lambda$ versus bath exponent $s$. We find $\lambda=\nu\simeq 1/s$ as $s\ra 0^-$, as found for $s\ra 0^+$ (see Fig.\ \ref{expon_nu}).

\subsubsection{Dynamical susceptibility}
\label{sectionds}
The local dynamical susceptibility $\cl(\w)$ can be computed using the NRG via its imaginary part
\be
\cl''(\w)=\frac{\pi}{Z} \sum_{m,n}|\bra n| S_z|m\ket|^2 \left(\mbox{e}^{-\beta E_n}-\mbox{e}^{-\beta E_m}\right)\delta(\w-E_{nm}),
\label{dspoles}
\ee
with $\cl''(\w)=\cl''(-\w)$, $E_{nm}=E_n-E_m$, and $Z=\sum_m e^{-\beta E_m}$ being the partition function.    For $T=0$,
\be
\cl''(\w)=\frac{\pi}{Z_0} \sum_{n}|\bra n| S_z|0\ket|^2 \delta(|\w|-E_{n0})~\mbox{sgn}(\w) 
\label{dspoleszero}
\ee
with an implicit summation over ground states $|0\ket$ in cases where the degeneracy $Z_0>1$.
Because only a small fraction of states are kept from one NRG iteration to the next, it is iteration $N$ that contains the most appropriate information for dynamics at energy scales $D\Lambda^{-N/2}$. Thus  Eq.\ (\ref{dspoles}) is evaluated at each NRG iteration for a frequency window spanning  $\w_N=\alpha D \Lambda^{-N/2}$.
The result is a discrete set of delta-function peaks that must be broadened to recover a continuous spectrum.  We follow 
 well-established procedure,\cite{Sakai:89} applying Gaussian
 broadening on a logarithmic scale, viz
\be
\delta(\w-\w_n)\ra\frac{\mbox{e}^{-b^2/4}}{\sqrt{\pi}b\w_n}\mbox{exp}\left[-\frac{(\ln\w-\ln\w_n)^2}{b^2}\right],
\label{broadening}
\ee
taking the broadening parameter $0.3<b<0.5$. 
The static local susceptibility (considered in Sec.\ \ref{statics}) follows via Hilbert transformation as
\be
\cl(T; \w=0)=\int_{-\infty}^{\infty}\frac{\mbox{d}\epsilon}{\pi}\ \frac{\cl''(\epsilon, T)}{\epsilon}.
\ee

\begin{figure}
\begin{center}
\epsfig{file=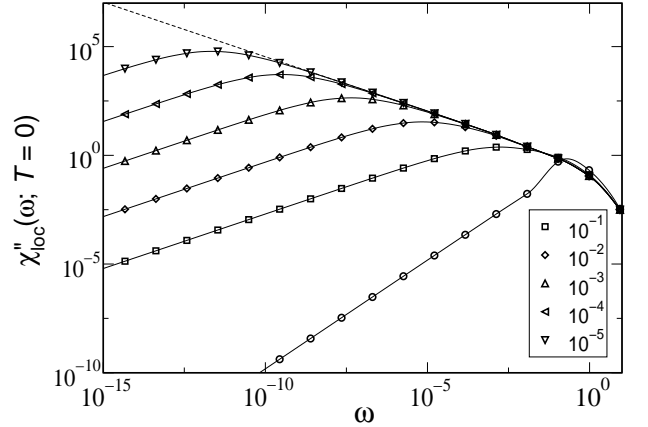, width=7cm, angle=270}
\end{center}
\caption{Imaginary part of the local dynamical susceptibility $\cl''(\w; T=0)$ for bosonic bath exponent $s=0.5$, Kondo coupling $J=0.5$ and a range of $g<g_c\simeq 1.216$ in the Kondo-screened phase.  The circles correspond to $g=0$ while the legend specifies $g_c-g$ for the remaining data. As $g\ra g_c^-$ the curves follow the quantum-critical form for $T_*\ll \w \ll \tkn$. The dashed line is a fit to the latter [Eq.\ (\protect\ref{dscrit})] which yields an exponent $y=s$ to within less than 0.5\%. See text for further comments.}
\label{dynsus}
\end{figure}

\textit{Zero temperature.} --- Figure \ref{dynsus} shows $\cl''(\w)$ at $T=0$ on a logarithmic scale for a sequence of increasing couplings $g<g_c$.  For zero coupling to the bosonic bath, the familiar Kondo-model result obtains: $\cl''(\w)\propto \w$ for $\w \ll \tkn$.   However,  for any nonzero $g$ in the Kondo phase, $\cl''(\w)\propto \w^s$ as $\w\ra 0$, corresponding to power-law relaxation $\cl(t)\propto t^{-(1+s)}$ for long times.  

\begin{figure}
\begin{center}
\epsfig{file=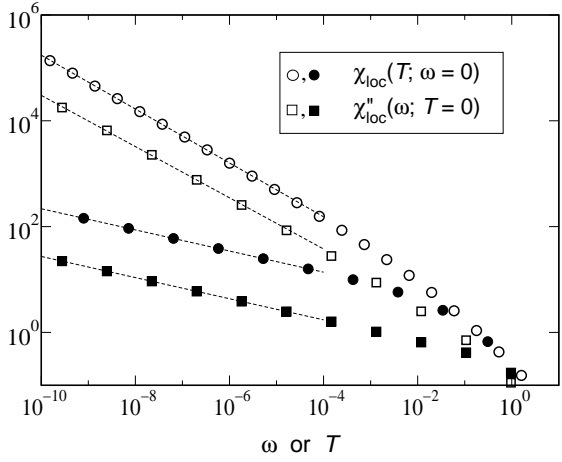, width=6.5cm, angle=270}
\end{center}
\caption{Critical $\cl''(\w;g=g_c, T=0)$ versus $\w$ [$\cl(T; g=g_c, \w=0)$ versus $T$]  for $J=0.5$ and $s=0.2$ (filled symbols) and $s=0.5$ (open symbols). It is seen clearly from the equality of the slopes that the exponent $y=x$.  Dashed lines are fits to the data and yield exponents $y=x=s$ to within the estimated numerical uncertainty. }
\label{xyexpon}
\end{figure}

At the QCP ($g=g_c$) we find the scale-free form
\be
\tkn\cl''(\w;g=g_c, T=0)\propto \left(\frac{\w}{\tkn}\right)^{-y} \mbox{sgn}(\w), 
\label{dscrit}
\ee
for $\w\ll \tkn$, with 
\be
y=x=s
\ee
obeyed for all $0<s<1$ to within our estimated numerical error, which is typically less than 1\%.
This is illustrated in Fig.\ \ref{xyexpon}, where $\cl(T;\w=0)$ and $\cl''(\w;T=0)$ are plotted on a logarithmic scale for bath exponents $s=0.2$ and $s=0.8$, and $g=g_c$ in either case.  The result $x=y$ is reflected in the equality of the slopes of the fits (shown as dashed lines).
  
\begin{figure}
\begin{center}
\epsfig{file=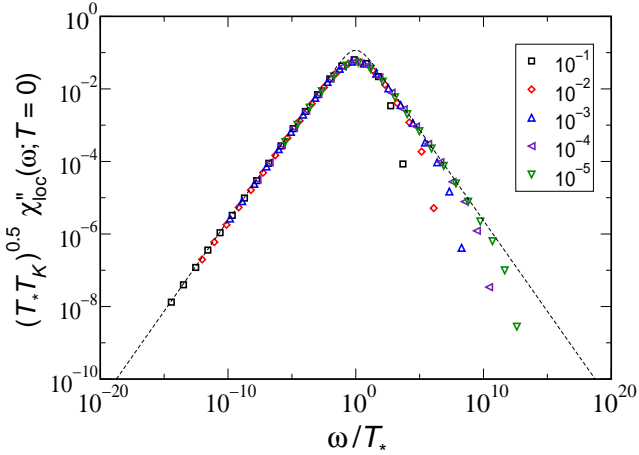, width=7cm, angle=270}
\end{center}
\caption{(Color online) Scaling of the $T=0$ dynamical suscceptibility for s=0.5 and $g\ra g_c^-$ --- see Eq.\ (\protect\ref{dsscalform}). The data were obtained for $J=0.5$.  The legend specifies $g_c-g$ and $T_*$ is here defined as the location of the peak maximum in Fig.\ \protect\ref{dynsus}, marking the crossover from quantum-critical $\w^{-s}$ behavior for $\w\gg T_*$  to $\w^s$ for $\w\ll T_*$.  For the data presented, $T_*$ varies over ten orders of magnitude.  Excellent data collapse is observed up to scales set by the bare Kondo temperature $\tkn/T_*$.  The dashed line is a simple fit taking $\phi_{s,K}(z)\propto z^{s}/(1+z^{2s})$, which does remarkably well at capturing the entire scaling curve.}
\label{scaldynsus}
\end{figure}

For small deviations from $g=g_c$ the critical power law behavior is cut off below a crossover scale $T^*$ that vanishes at the QCP according to Eq.\ (\ref{nudef}).  Thus, the correlation length exponent $\nu$ can also be extracted from $\cl''(\w;T=0)$ and agrees with the results of Fig.\ \ref{expon_nu} to within our estimated numerical error.
In the vicinity of the QCP we find the zero-temperature scaling form 
\be
\tkn\cl''(\w; T=0)=\left(\frac{T_*}{\tkn}\right)^{-s}\phi_{s,p}\left(\frac{\w}{T_*}\right)
\label{dsscalform}
\ee
for $T_*\ll \tkn$, where $p=K$ or $L$ denotes the phase.  For $z\gg 1$, $\phi_{s,p}(z)=z^{-s}$ for either phase and thus $\cl''(\w)$ is independent of $T_*$ for $\w\gg T_*$ (see, e.g., Fig.\ \ref{dynsus}), following the quantum-critical form $\tkn\cl(\w)\propto (\w/\tkn)^{-s}$ before ultimately being cut off by the nonuniversal scale $\tkn/T_*$.  Such scaling behavior is clearly illustrated in Fig.\ \ref{scaldynsus}, which shows scaling for $g\ra g_c^-$.

\begin{figure}
\begin{center}
\epsfig{file=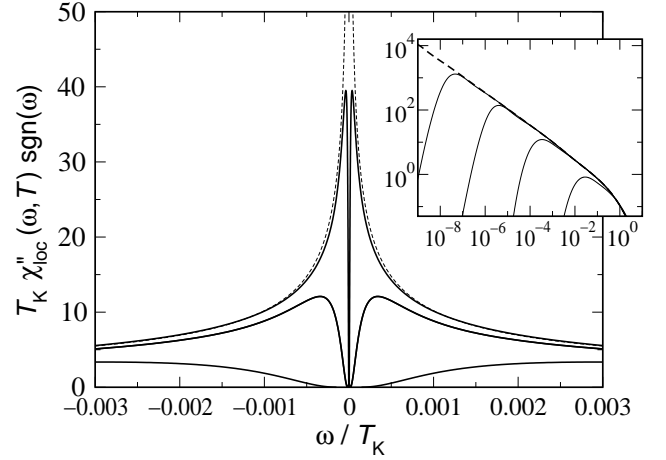, width=7cm, angle=270}
\end{center}
\caption{Imaginary part of the dynamical local susceptibility $\dcl(\w,T)$ versus $\w$, at the critical coupling $g=g_c$ for $s=0.5$ and a sequence of increasing temperatures $T/\tkn=0$ (dotted line), $10^{-5}$, $10^{-4}$ and $10^{-3}$.  Temperature is the single low-energy scale, setting the location of the peak maximum.  The inset shows data on a logarithmic scale for $T/\tkn=10^{-2}$, $10^{-4}$, $10^{-6}$ and $10^{-8}$. The $T=0$ data are shown as a dashed line and follow the critical power law Eq.\ (\protect \ref{dscrit}) with $y=0.5$.}
\label{dsvsT1}
\end{figure}

\begin{figure}
\begin{center}
\epsfig{file=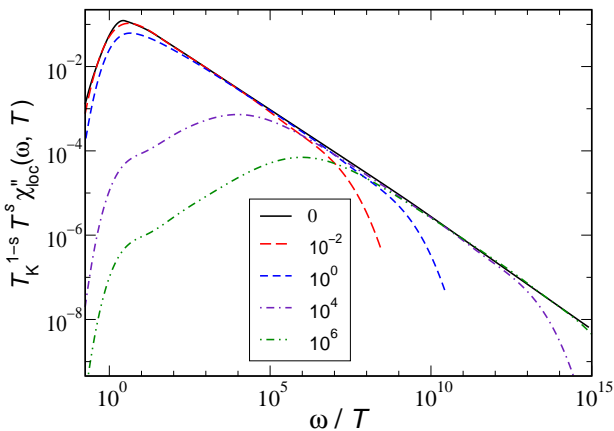, width=7cm, angle=270}
\end{center}
\caption{(Color online) $\w/T$ scaling of the dynamical local susceptibility in the Kondo phase for $s=0.5$ and values of $T_*/T$ specified in the legend.  The QCP spectrum obtains for $T_*/T=0$.  The tails of the curves fall onto  the QCP form for $T_*/T\ll \w/T\ll \tkn/T$, as shown clearly by the $T_*/T=10^6$ data.  Here, $T_*$ is defined as the crossover scale in $\cl''(\w;T=0)$ from $\w^{-s}$ to $\w^s$ behavior.}
\label{wovertscal}
\end{figure}

\textit{Finite temperature.} --- The result $y=x$ implies that the relaxation rate is linear in temperature, characteristic of critical physics below the upper critical dimension.  Quantum-critical fluctuations are cut off on a timescale $\tau = \hbar/k_BT$.  Thus, for nonzero $T\ll\tkn$ the quantum-critical form  Eq.\ (\ref{dscrit}) is cut off for frequencies $\w\approx T$, as is shown clearly in the inset to Fig.\ \ref{dsvsT1}.  For $\w/T\ll \tkn/T$ this leads 
 to $\w/T$ scaling of the form
\be
\tkn\cl''(\w,T;g=g_c)=\left(\frac{T}{\tkn}\right)^{-s}\phi_s\left(\frac{\w}{T}\right).
\label{wotform}
\ee
with $\phi_s(z)=z^{-s}$ for $z\gg 1$.  The numerically obtained QCP scaling curve is shown in Fig.\ \ref{wovertscal} as the thick line.  Though the NRG method is unreliable for $\w\lesssim T$ (Refs.\ \onlinecite{Costi:94} and \onlinecite{Ingersent:02}) due to truncation error, we find a consistent description of the finite-$T$ scaling properties of the model as discussed further below.  See also Fig.\ 3 of Ref.\ \onlinecite{Glossop:05a}, which demonstrates $\w/T$ scaling for $s=0.2$ and different values of $\tkn$.

\label{clomega}

For any finite $T_*$, the local dynamical susceptibility obeys the full scaling form to be expected in the vicinity of an interacting QCP\cite{Vojta:04a}:
\be 
\tkn\cl''(\w,T)=\left(\frac{T}{\tkn}\right)^{-s}\Phi_{s,p}\left(\frac{\w}{T},\frac{T_*}{T}\right).
\ee
 Figure \ref{wovertscal} plots $\w/T$ scaling curves obtained via NRG calculations for several values of the ratio $T_*/T$ in the Kondo-screened phase of the model for $s=0.5$. 
For $\w/T\gg T_*/T$ the curves follow the QCP ($T_*=0$) form [Eq.\ (\ref{wotform})], which is of course independent of phase, $p=K$ or $L$.  Clearly, as $T_*/T\ra 0$ the curves follow the QCP result more closely, i.e., $\phi_s(z)=\Phi_{s,p}(z,0)$.   For the opposite extreme $T_*/T \ra \infty$, the $T=0$ result obtains when suitably rescaled (see Fig.\ \ref{scaldynsus}), such that $\Phi_{s,p}(z',z)=z^{-s}\phi_{s,p}(z'/z)$ for $z,z'\ra \infty$.   Since all data were obtained for finite $\tkn$, the tail behavior is ultimately cut off when $\w/T$ increases to scales of order $\tkn/T$, as seen in the figure.
The $T_*/T$ dependence of the $\w/T$ scaling spectrum  has also been discussed in the context of the Kondo model with a pseudogap host density of states $\rho(\epsilon)\propto |\epsilon|^r$, though  the QCP of the latter model belongs to a different universality class. \cite{Glossop:05b,Fritz:06}

To summarize this subsection, we have studied in detail the local magnetic response for the Ising-anisotropic BFKM, extracting a range of critical exponents.  For $0<s<1$ we find that the exponents satisfy hyperscaling relations and $\w/T$-scaling characteristic of an interacting critical fixed point. All exponents associated with the local magnetic susceptibility can be determined by knowledge of just $x$ and $\nu$.
   Furthermore, we find excellent agreement with corresponding results for the spin-boson model, confirming that the two models belong to the same universality class.  As pointed out in Ref.\ \onlinecite{Vojta:04}, these results demonstrate the failure of a mapping\cite{Fisher:72,Kosterlitz:76} of the SBM to a one-dimensional classical Ising model with long-range interactions falling off as $1/r^{1+s}$.
  The latter model is above its upper critical dimension for $0<s<\frac{1}{2}$ and the exponents  there  take mean-field values with hyperscaling relations violated.  However, in the range $\frac{1}{2}<s<1$ the SBM and classical long-range Ising model appear to be equivalent.  See Refs.\ \onlinecite{Vojta:04} and \onlinecite{Vojta:04a} for further discussion of this issue.

\subsection{Spectral function}
\label{secspec}
We now turn to a discussion of the local single-particle spectrum, paying particular attention to the destruction of the Kondo resonance at the quantum-critical point.
In this section we calculate the impurity spectral function 
\begin{equation}
A(\w)=\frac{1}{Z} \sum_{n,m} |\bra n|d_{\sigma}^{\dagger}|m\ket|^2 \delta(\w-E_{n,m}) \left
(\mbox{e}^{-\beta E_m}+\mbox{e}^{-\beta E_n}\right)
\label{deltafns}
\end{equation}
for the  Ising-symmetry Bose-Fermi Anderson model described by
\bea
\hat{H}_{\mbox{\ssz{int}}}&=&\sum_{\sigma}(\epsilon_{d}+\mbox{$\frac{1}{2}$}U\hat{n}_{d
-\sigma})\hat{n}_{d\sigma}
+V \sum_{\kk,\sigma}(d^{\dagger}_{\sigma}c_{\kk\sigma}+\mbox{H.c.})\nonumber\\
&+& \frac{1}{2}g(\hat{n}_{d\uparrow}-\hat{n}_{d\downarrow})\sum_{\qq}(\phi_{\qq} + \phi^{\dagger}
_{-\qq}).
\label{anderham}
 \eea
Here, $\epsilon_{\mbox{\ssz{d}}}$ is the impurity level energy, $U>0$ is the on-site Coulomb interaction, and $V$ is the hybridization between impurity and conduction band.  In the limit where charge fluctuations at the impurity site are supressed, the Anderson model maps onto the Kondo model under a Schrieffer-Wolff transformation. \cite{Schrieffer:66}  For particle-hole-symmetric impurity parameters $\epsilon_d=-\frac{1}{2}U$, the mapping takes $8\Gamma_0/\pi U \ra \rho_0J_0$ with $\Gamma_0\equiv\pi V^2 \rho_0$ the hybridization width.

 The NRG approach described above, with the standard broadening procedure Eq.\ (\ref{broadening}) applied to Eq.\ (\ref{deltafns}), also leads also to a satisfactory description of $A(\w)$ in the Bose-Fermi case, where 
the quantum phase transition is manifest as a collapse of the central Kondo resonance.
All results for $A(\w)$ were obtained for NRG discretization parameter $\Lambda=3$, retaining $N_s=800$ states from one iteration to the next, and taking bosonic truncation parameter $N_b=8$.  Again, the half-bandwidth $D$ is taken to be the energy unit unless  stated otherwise.  
\begin{figure}
\begin{center}
\epsfig{file=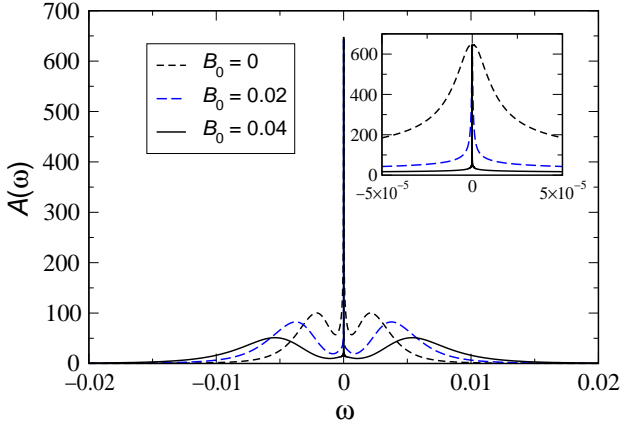, width=6.5cm, angle=270}
\end{center}
\caption{(Color online). Single-particle spectrum $A(\w)$ versus $\w$ for $s=0.8$, $U=-2\epsilon_d=0.005$, $\Gamma_0=5\times 10^{-4}$ and Kondo phase bosonic couplings $B_0<B_{0,c}\simeq 0.04704$ specified in the legend.  The inset shows the central resonance on an expanded scale.  See text for discussion.}
\label{spec1}
\end{figure}
 We use broadening parameter $b=0.3$, which gives the most accurate recovery of the standard $B_0=(K_0g_0)^2=0$ Fermi-liquid-theory result $A(\w=0)=1/\pi\Gamma_0$ (see, e.g., Ref.\ \onlinecite{Hewson:93}), which is typically satisfied to within 2\% within NRG. \cite{Costi:94}  We do not vary $b$ for $B_0>0$.

\begin{figure}
\begin{center}
\epsfig{file=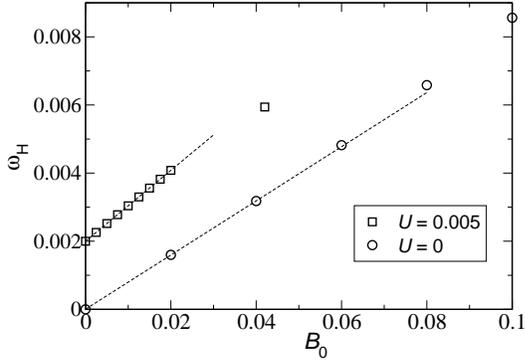, width=6cm, angle=270}
\end{center}
\caption{Location of Hubbard band maximum $\w_{\mbox{\ssz{H}}}$ as a function of bosonic bath coupling $B_0$ for $s=0.8$, $\Gamma_0=5\times 10^{-4}$, and the $U=-2\epsilon_d$ values shown.
The dashed lines are linear fits to the data for small $B_0$.}
\label{hubbard}
\end{figure}

Fig.\ \ref{spec1} compares the single-particle spectrum $A(\w)$ --- obtained for impurity parameters $U=-2\epsilon_d=0.005$ [such that $A(\w)=A(-\w)$] and hybridization $\Gamma_0=5\times 10^{-4}$ --- for $B_0=0$ and two couplings $0<B_0<B_{0,c}\simeq 0.047043(1)$.  The coupling to the bosonic bath has two main effects. First, the Hubbard satellite bands  are displaced to higher energies from their $B_0=0$  positions, which in the Kondo limit ($J \ll 1$) are given by $\w_{\mbox{\ssz{H}}}\simeq\pm\frac{1}{2}U$.  The shift is linear in $B_0$ for small $B_0$, as is demonstrated in Fig.\ \ref{hubbard}.
Second, while the central Kondo resonance remains pinned to the $B_0=0$ Fermi-liquid-theory result $A(\w=0)=1/\pi\Gamma_0$, which remains satisfied to within 2\%,  its halfwidth is significantly reduced, as can be seen  in the inset to Fig.\ \ref{spec1}.

\begin{figure}
\begin{center}
\epsfig{file=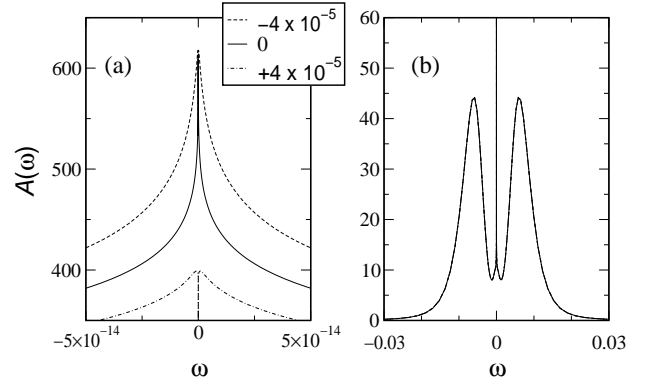, width=6.5cm, angle=270}
\end{center}
\caption{Single-particle spectrum vs $\w$ for bath exponent $s=0.8$,
impurity parameters $U=-2\epsilon_d=0.005$,
hybridization strength $\Gamma_0=5 \times 10^{-4}$, and discretization
$\Lambda=3$.
Shown for $B_0-B_{0,c} = 0$ (dashed line),$\ 4\times \pm 10^{-5}$ on (a) the scale set by the Kondo resonance; and (b) the scale of the on-site Coulomb repulsion $U$.  The critical coupling $B_{0,c}\simeq 4.704\times 10^{-2}$.}
\label{spec}
\end{figure}

At the critical coupling $B_0=B_{0,c}$, we find that the  pinning of $A(0)$ persists and the resonance has nonzero width.  This is illustrated in Fig.\ \ref{spec}, which shows three spectra: one at the QCP and one close to it in either phase.
 Figure\ \ref{spec}(a) plots the spectra on the scale of the central feature and shows clearly the pinning of the critical spectrum (solid line).  For $B_0>B_{0,c}$ the Kondo resonance collapses leaving a double peak structure with characteristic scale $T_*$ that vanishes as $B_0\ra B_{0,c}^+$.     However, contrary to expectations, we find that $A(0)$ is nonzero and decreases with increasing $B_0$.  For fixed $B_0>B_{0,c}$ we find that $A(0)$ does not significantly vary with $N_b$ in the range $8\le N_b \le 16$.   Though the three spectra in Fig.\ \ref{spec} show very different behaviors at low frequencies, on the scale of the Hubbard satellite bands [see Fig.\ \ref{spec}(b)] they are indistinguishable. 

In the Kondo phase, the scale $T_*$, which vanishes as $B_0\ra B_{0,c}^-$, is manifest in $A(\w)$ as the crossover scale to the Fermi liquid form $A(0)-A(\w)\propto \w^2$ as $\w\ra 0$.

\begin{figure}
\begin{center}
\epsfig{file=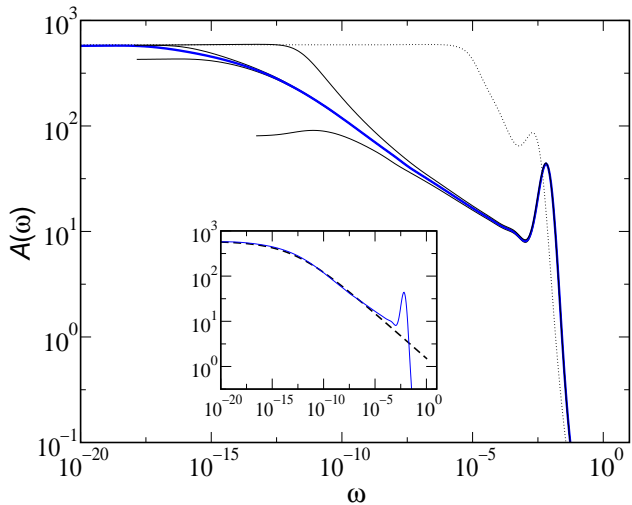, width=7cm, angle=270}
\end{center}
\caption{(Color online) Single-particle spectrum $A(\w)$ versus $\w$, shown on a logarithmic scale, for $s=0.8$, $U=-2\epsilon_d=0.005$, $\Gamma_0=5\times 10^{-4}$, and bosonic couplings $B_0=0$ (dotted line), $B_{0,c}$ (thick solid line), $B_{0,c}\pm 10^{-5}$ and $B_{0,c}\pm 10^{-3}$.  See text for discussion.  The inset shows a fit to the QCP spectrum of form $A(\w)=[\pi\Gamma_0(1+k|\w|^a)]^{-1}$ with $a=0.2$, which is followed well over several orders of magnitude until it is cut off by the Hubbard satellite bands.}
\label{logspec}
\end{figure}

Figure \ref{logspec} shows $A(\w)$ for $B_0=0$ (dotted line) and a set of $B_0$ values close to the critical coupling; $A(\w)$ for $B_0=B_{0,c}$ is shown as a thick solid line.  
The crossover scale $T_*$, at which the Fermi liquid ``shoulder'' is observed  vanishes as $B_0\ra B_{0,c}^-$.  On the localized side, the low frequency peak vanishes as $B_0\ra B_{0,c}^+$.  Thus, regardless of phase, the critical spectrum is followed to progressively lower frequencies as the transition is approached.  The correlation length exponent $\nu(s)$ can be calculated from the crossovers in $A(\w)$ described, and agrees well with the values determined for the BFKM in Sec.\ \ref{secresults}.

In the quantum-critical region $|\w|\gg T_*$, $A(\w)$ is of form
\be
A(\w=0)-A(\w)\propto |\w|^a,
\ee
with $A(\w=0)=1/\pi\Gamma_0$ satisfied for all $0<s<1$ to within 2\%.
To extract the power $a$ we fit the critical spectrum ($B_0=B_{0,c}$) to the form
\be
A(\w)=[\pi\Gamma_0(1+k|\w|^a)]^{-1},
\label{specfit}
\ee
shown as the thick dashed line in the inset to Fig.\ \ref{logspec} for $s=0.8$.  The fit is followed well over many orders of magnitude and is ultimately cut off by the Hubbard satellite bands.  Though there remains a small numerical uncertainty in the fits, largely due to the fact that the zero-frequency pinning is only satisfied to within 2\%, our results suggest that $a=1-s$ for all $0<s<1$ with $k$ a constant depending on $s$ and $\tkn$.

\begin{figure}
\begin{center}
\epsfig{file=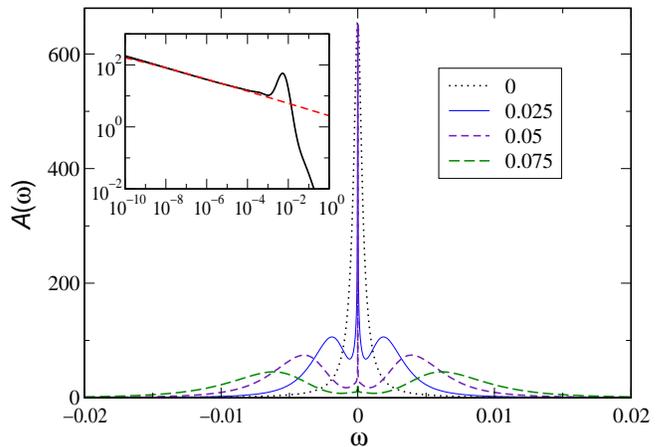, width=6.75cm, angle=270}
\end{center}
\caption{(Color online) Single-particle spectrum $A(\w)$ versus $\w$ for non-interacting limit $U=0$ with $\Gamma_0=5\times 10^{-4}$, $s=0.8$, and the $B_0$ values shown in the legend.  With increasing $B_0<B_{0,c}\approx 0.0648$ the central Lorentzian peak narrows rapidly, developing shoulders that separate from the parent peak to form distinct bands.  In the quantum-critical regime ($|\w|\gg T_*$) we find the same frequency dependence as for $U>0$; the dashed line in the inset shows a fit of form Eq.\ (\protect\ref{specfit}) with $a=0.2$ to data for $B_0=0.0648$ (K phase) and $0.0650$ (L phase). As for $U>0$ the localized phase spectrum contains a double peak structure about $\w=0$ with $A(\w=0)>0$.  See text for further comment.}
\label{specU0}
\end{figure}

Surprisingly, the key features described above are all present in the \textit{noninteracting} limit ($U=\epsilon_d=0$) of the Anderson model, Eq.\ (\ref{anderham}).  Figure \ref{specU0} shows the spectral function $A(\w)$ obtained for $s=0.8$ with $U=0$, fixed $\Gamma_0=5\times 10^{-4}$, and a sequence of increasing bosonic couplings $B_0$. For $B_0=0$ (dotted line) the model is exactly solvable and it is well known that $A(\w)$ has a simple Lorentzian form $A(\w)=(\Gamma_0/\pi)/(\w^2+\Gamma_0^2)$, which is recovered well by the NRG broadening procedure.  With increasing $B_0<B_{0,c}\approx 0.0648$ the central peak narrows rapidly, developing symmetric shoulders that separate into distinct bands redolent of the Hubbard satellites for the regular symmetric Anderson model ($B_0=0$) for $U=-2\epsilon_d\gg 1$.  As found for $U>0$, the position of the satellite bands increases linearly in $B_0$ for small $B_0$, as shown in Fig.\ \ref{hubbard}.  At the critical coupling the central resonance has nonzero width and remains pinned at $A(0)=1/\pi\Gamma_0$.  It is again possible to fit the critical spectrum to the form Eq.\ (\ref{specfit}) and, as for $U=0.005$ (Fig.\ \ref{logspec}), we find $a=0.2$ for $s=0.8$.  This is demonstrated in the inset to Fig.\ \ref{specU0}, which shows data for $B_0=0.0648$ (Kondo phase) and $0.0650$ (localized phase).  In the frequency range shown the two data sets coincide --- a reflection of the fact that $|\w|\gg T_*$ --- and are well described by the fit discussed above, which is plotted as a dashed line.

\begin{figure}
\begin{center}
\epsfig{file=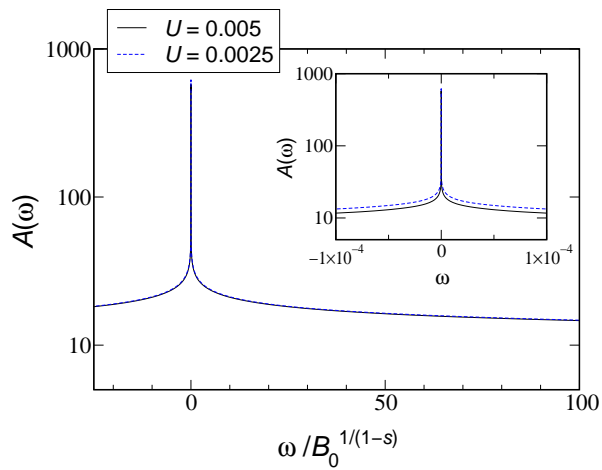, width=6.5cm, angle=270}
\end{center}
\caption{(Color online) Critical single-particle spectrum $A(\w)$ for
$s=0.8$, $U=0.005$ and $0.0025$, $\epsilon_d=-\frac{1}{2}U$, $\Gamma_0=5\times 10^{-4}$, and $B_0=B_{0,c}$ in either case.  The main figure shows scaling in terms of $B_0^{1/(1-s)}\propto \tkn$, while the inset shows the two data sets on an absolute scale.}
\label{woverb0scal}
\end{figure}

At the quantum-critical point itself we find that spectra scale according to
\be
\pi\Gamma_0A(\w)=\psi_s(\w/B_0^{\frac{1}{1-s}}),
\label{specscalb0c}
\ee
as demonstrated in Fig.\ \ref{woverb0scal}.  In the vicinity of the quantum-critical point single-particle spectra scale according to
\be
\pi\Gamma_0A(\w)= \psi_{s,p}\left(\frac{\w}{T_*},\frac{\tkn}{T_*}\right)
\ee
where $p=$ K or L denotes the phase.  For the QCP, where $T_*=0$, and $B_{0,c}\propto \tkn^{1-s}$, the form reduces to that of Eq.\ (\ref{specscalb0c}), which scaling is demonstrated explicitly in Fig.\ \ref{woverb0scal}.

\seceq
\section{Results: Ohmic dissipation, $s=1$}
\label{s1section}
For any fixed $\tkn$, the line of continuous quantum phase transitions in the $g$-$s$ plane terminates at a Kosterlitz-Thouless-like transition at $s=1$. 
As is to be expected from the well-known mapping discussed in Sec.\ \ref{mapping}, the key physics of the BFKM with Ohmic dissipation is already present in the anisotropic Kondo model.  The Ohmic BFKM corresponds to an Ohmic spin-boson model with dissipation strength $\alpha$ containing contributions from the Kondo term in the Hamiltonian, Eq.\ (\ref{akmham}) with $J_{\perp}=J_z=J_0$, \textit{and} from $B_0$, Eq.\ (\ref{bosebath}).  For $B_0=0$ and $0<\rho_0J_0 \ll 1$, the resulting SBM lies in the delocalized phase with $\alpha=1^-$;  an exponentially-small renormalized tunelling amplitude, corresponding to the Kondo scale, sets the scale for the low-energy physics.  By increasing $B_0>0$, the sytem may be further driven towards a Kosterlitz-Thouless transition at $\alpha=1$, where the renormalized tunelling amplitude vanishes, and eventually into the localized phase $\alpha>1$.  This is, of course, the essential physical behavior of the \textit{anisotropic} Kondo model.  In other words, the principal effect of coupling to an Ohmic bosonic bath is to introduce anisotropy into the corresponding Kondo model: $\rho_0 J_{\perp}-\rho_0 J_{z}\propto B_0$.

In this section we present results of direct calculations for the Bose-Fermi Kondo and Anderson models obtained using the NRG approach described in Sec.\ \ref{secnrg}.  We begin by discussing the Kosterlitz-Thouless nature of the NRG flow of effective couplings.   We then cast results for the static local susceptibility in terms of a noisy quantum box,\cite{LeHur:04} showing that perfect Coulomb-staircase behavior obtains above a critical coupling to the Ohmic bath.  We conclude by discussing the single-particle spectral function for the Bose-Fermi Anderson model with Ohmic dissipation, along with its universal scaling properties.

\begin{figure}
\begin{center}
\epsfig{file=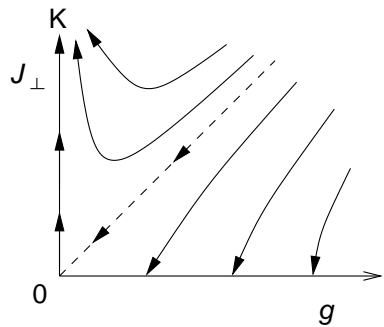, width=5cm}
\end{center}
\caption{Schematic of NRG flow of effective couplings for the BFKM with Ohmic dissipation.  The dashed line represents the phase boundary in the $J$-$g$ plane.  A crossover scale $T_*$ vanishes upon approaching the transition from the Kondo-screened phase.  See text for discussion.}
\label{flow1}
\end{figure}

\begin{figure}
\begin{center}
\epsfig{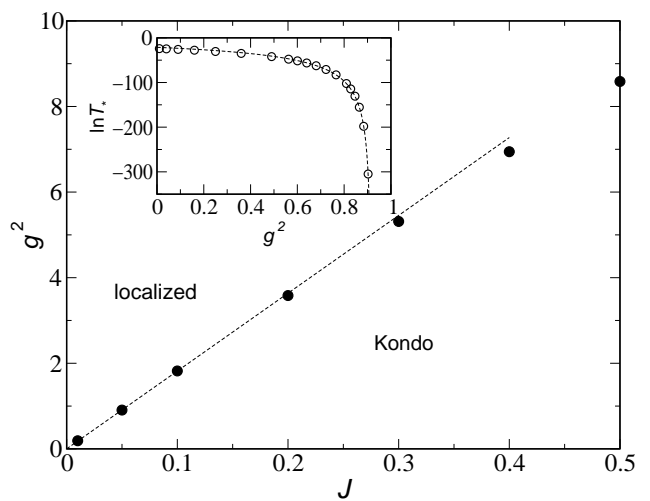}
\end{center}
\caption{Phase boundary between Kondo ($g<g_c$) and localized phases ($g>g_c$) as a function of Kondo exchange interaction $J$ for Ohmic dissipation. For $J\ll 1$, $g_c^2\propto J$ (dashed line).  The inset shows the vanishing of the crossover scale $T_*$, with the latter obtained from NRG levels flows, as the phase boundary is approached from the Kondo phase for $J=0.05$.  The dashed line shows a fit to the form Eq.\ (\protect\ref{bfkms1scale}) with $g_c\simeq 0.958$.}
\label{s1pb}
\end{figure}

For $s\ra 1^-$ the critical fixed point merges with the weak-coupling fixed point located at $J=g=0$, as found in Ref.\ \onlinecite{Zhu:02}.  A schematic NRG flow diagram for $s=1$ is given in Fig.\ \ref{flow1}.  The phase boundary between Kondo-screened and localized phases is shown as a dashed line. Figure \ref{s1pb} shows the numerically obtained phase boundary for the Ohmic case.  As expected by analogy to the AKM, $g_c^2\propto J$ for $J\ll 1$. For $g\ra g_c^-$ the NRG level flow is first towards the localized fixed point at $J=g=0$.  At a low-energy scale $T_*$ (large NRG iteration number $N_*$), the NRG levels cross over  to their stable Kondo-fixed-point values for $s=1$. In contrast to the sub-Ohmic case $0<s<1$, there is no crossover scale observed when approaching  the transition from the localized side, corresponding to a line of stable fixed points being reached for $g>g_c$.

\begin{figure}
\begin{center}
\epsfig{file=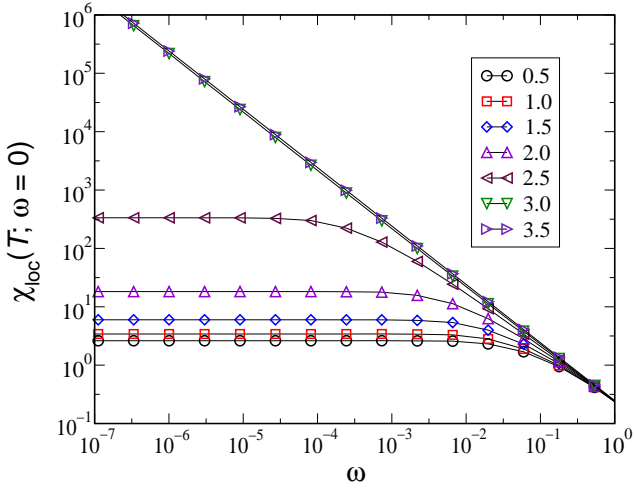, width=6.75cm, angle=270}
\end{center}
\caption{(Color online) $\cl(T;\w=0)$ versus $T$ for the Ohmic case $s=1$, $J=0.5$ and bosonic couplings shown; $g_c\simeq 2.92$.  On the Kondo side, a clear crossover scale exists, equivalent to the renormalized Kondo scale $\cl^{-1}(0,0)$, and vanishes continuously as $g\ra g_c^-$.  No such scale is present for the localized phase.
 Data obtained for $\Lambda=9$, $N_b=12$, and $N_s=800$.  See text for further comment.}
\label{chilocs1}
\end{figure}

The above comments are well illustrated by Fig.\ \ref{chilocs1}, which shows $\cl(T; \w=0)$ for $s=1$, $J=0.5$ and a range of dissipation strengths $g$ spanning $g_c\simeq 2.92$.  The observed behavior should be compared to that of Fig.\ \ref{chilocvst} for $s=0.5$, which is typical of the sub-Ohmic case $s<1$.  For Ohmic dissipation and $g<g_c$, the impurity spin is quenched below a crossover scale $T_*$ that vanishes continuously as $g\ra g_c^-$.  
\begin{figure}
\begin{center}
\epsfig{file=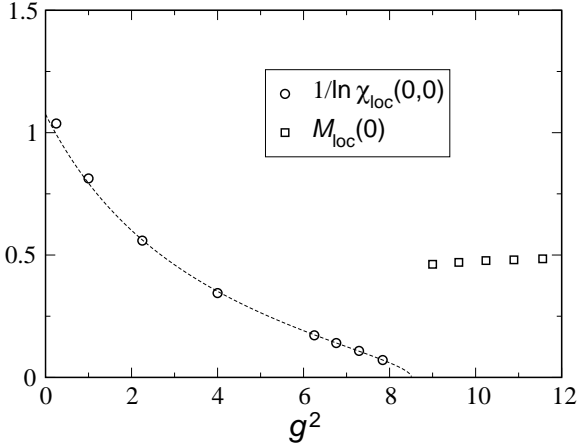, width=6.5cm, angle=270}
\end{center}
\caption{Behavior of the renormalized Kondo scale $\cl^{-1}(T=0,\w=0)$ ($\propto T_*$ for $s=1$) and the order parameter $M_{\mbox{\ssz{loc}}}(T=0)$ for the Ohmic case $s=1$ and $J=0.5$.  The dashed line is a fit of form Eq.\ (\ref{bfkms1scale}).  In the Ohmic case the order paramter undergoes a clear jump at the transition point $g=g_c\simeq 2.92$.  See text for further discussion.}
\label{s1scale}
\end{figure}
The renormalized Kondo scale $\cl(0,0)^{-1}$ ($\propto T_*$ for $s=1$) likewise vanishes and we find
\be
\ln~T_*=a_0-\frac{a_1}{\sqrt{g_c^2-g^2}}.
\label{bfkms1scale}
\ee 
Such a form follows from a poor-man's scaling analysis for the anisotropic Kondo model,\cite{Anderson:70} which can be shown to yield
\be
\tkn=D\exp\left(1+\frac{1}{|\rho_0J_{z}|}-\frac{\pi}{\delta}\right),
\label{akmscale}
\ee
which holds for $0\ll\delta \ll -\rho_0J_z$ close to the transition with $\delta^2=(\rho_0J_{\perp})^2-(\rho_0J_{z})^2$, and recognizing that $g_c^2-g^2\propto (\rho_0J_{\perp}-\rho_0|J_z|)\simeq\delta^2/2J$.  A fit to the behavior Eq.\ (\ref{bfkms1scale}) is shown in Fig.\ \ref{s1scale}, yielding a value for $g_c\simeq 2.92$  in good agreement with that obtained via NRG level flows.  The corresponding fit for $J=0.05$ is shown in the inset to Fig.\ \ref{s1pb}.   The constants $a_0$ and $a_1$ vary as $1/\sqrt{J}$ for $J\ll 1$, as expected from Eq.\ (\ref{akmscale}).

 For $T\gg T_*$, we find $\cl(T;\w=0)\propto T^{-1}$ with logarithmic corrections. For $g>g_c$, $\cl(T;\w=0)\propto 1/T$ shows Curie-like behavior characteristic of a freely fluctuating spin.  However, as $g\ra g_c^+$, the order parameter $M_{\mbox{\ssz{loc}}}(0)$ does not vanish continously, but rather jumps to zero at $g=g_c$ with the size of the jump being roughly 90\% of the full saturation value.

\textit{The noisy quantum box.}---The case of Ohmic dissipation in the BFKM has recently been studied in the context of  gate-voltage fluctuations in a quantum box.
A quantum box consists of a small metallic grain (or dot) with a dense spectrum coupled to reservoir leads.  Gate electrodes control the electrostatic energy of the dot,
\be
E_Q=\frac{(Q-eN)^2}{2C},
\ee
where $N$ is proportial to the gate voltage, $Q$ is the charge on the dot and $C$ the capacitance.  Degeneracy points at half-integer values of $N$ (where $E_Q=E_{Q-e}$) lead to a unit jump in the number of electrons on the grain at that point, and thus to the familiar Coulomb-blockade ``staircase'' behavior for $\bra Q\ket$ versus $N$.  However, the sharp charging steps are rounded due to quantum fluctuations, as shown theoretically by Matveev \cite{Matveev:90} by mapping the problem to a two-channel Kondo model.  Le Hur has recently shown \cite{LeHur:04} that the supression of quantum charge fluctuations due to the dissipative effects of a fluctuating gate voltage can restore perfect Coulomb-staircase behavior.  This has been discussed in the context of a Bose-Fermi Kondo model, where the ``spin'' states of the impurity correspond to the two charge states (orbital pseudospin) of the grain close to a degeneracy point. Considering charging states $Q=0$ and $Q=1$ leads to the identification $\bra S_z\ket=\bra Q\ket-\frac{1}{2}$. The electrons themselves are taken to be spinless, thus assuming the presence of a strong in-plane magnetic field.  The gate voltage, which controls deviations away from the degeneracy point, i.e., differences in the energies of the spin states, plays the role of the local magnetic field $h$ in the impurity problem. The Kondo term represents tunneling from lead to grain and vice versa, while the bosonic bath, which is Ohmic, represents the voltage fluctuations.  It then follows that
\be
\frac{\partial \bra Q\ket}{\partial N}\propto \frac{\partial \bra S_z \ket}{\partial h}.
\ee
The static susceptibility $\cl(0,0)$ thus measures the slope of  $\bra Q\ket$ versus $N$ at the degeneracy point $N=\frac{1}{2}$. Our results indeed show a perfect restoration of the Coulomb staircase for large enough coupling to the bosonic environment.    
\begin{figure}for  $g\ge g_c\simeq 2.92$
\begin{center}
\epsfig{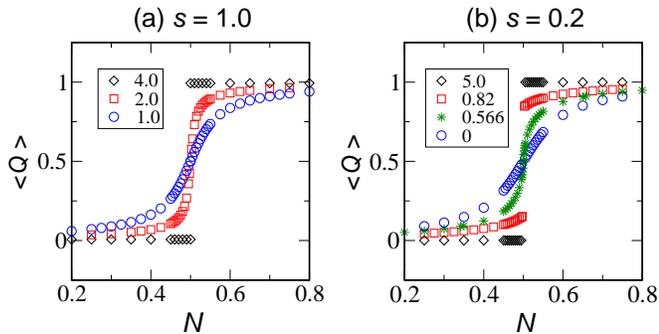}
\end{center}
\caption{(Color online) Average charge on dot $\bra Q\ket$ versus gate voltage $N$, for the noisy quantum dot system with (a) Ohmic and (b) sub-Ohmic dissipation.  The legends specify the dissipation strength $g$ in the BFKM.  For weak dissipation the curves $\bra Q\ket$ versus $N$ are linear through the degeneracy point $N=\frac{1}{2}$ due to Kondo screening of local moments.  In the Ohmic case  (a),  the curves undergo a large jump at $N=\frac{1}{2}$ for  $g\ge g_c\simeq 2.92$.  For $g=g_c\simeq 0.566$ in the sub-Ohmic case  (b), $\bra Q\ket-\frac{1}{2}$ vanishes in a power-law fashion at $N=\frac{1}{2}$, while for increasing $g>g_c$ it exhibits a jump of a magnitude that increases continuously from zero.}
\label{qvsn}
\end{figure}
Figure \ref{qvsn}(a) shows $\bra Q\ket$ versus $N$ plots for the Ohmic case where perfect Coulomb staircase behavior is restored for $g \ge g_c$. That there is a jump in $\bra Q\ket$ for $g=g_c$ itself stems from the Kosterlitz-Thouless nature of the transition discussed above.
For sub-Ohmic baths, by contrast, a continuous critical curve obtains at $g=g_c$, with a slope that is infinite at $N=\frac{1}{2}$.  For increasing $g>g_c$, the size of the jump in $\bra Q \ket$ at $N=\frac{1}{2}$ increases from zero, consistent with the behavior of $M(0)$ in Fig.\ \ref{s1scale}.
Our NRG results, which solve the Ohmic BFKM directly, are consistent with the picture described by perturbative RG \cite{LeHur:04} and by a bosonic NRG treatment following the mapping of the BFKM to a spin-boson model.\cite{Li:05}

\begin{figure}
\begin{center}
\epsfig{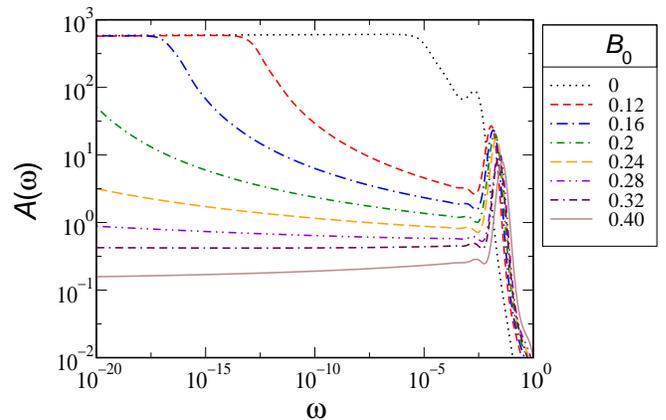}
\end{center}
\caption{(Color online) Single particle spectrum $A(\w)$ versus $\w$ for the Ohmic case $s=1$ and $U=-2\epsilon_d=0.005$, $\Gamma_0=5\times 10^{-4}$, and the $B_0$ values shown. See text for comments. }
\label{s1spec}
\end{figure}

We now turn to a discussion of  spectral functions for the Bose-Fermi Anderson model with Ohmic dissipation. Figure \ref{s1spec} shows spectra for  fixed $U=-2\epsilon_d=0.005$ and $\Gamma_0=5\times 10^{-4}$, for which the corresponding $J\simeq 0.25$, and a sequence of increasing dissipation strengths $B_0$. 
There are several key differences compared to the sub-Ohmic case.  First, and most obviously, the spectrum at the critical coupling $B_0=B_{0,c}\simeq 0.33$ is not pinned to the Kondo phase value $A(\w=0)=(\pi\Gamma_0)^{-1}$, but is much reduced.   Furthermore, there is essentially no frequency dependence to the critical spectral function below scales on the order of the Hubbard satellite bands.   For the parameters considered in Fig.\ \ref{s1spec}, $A(\w=0)\simeq 0.5$ at the critical coupling.  This form is followed by the $B_0=0.32$ data down to $T_*\simeq 10^{-120}$ which sets the scale for crossover to Kondo-phase behavior.

Second, the vanishing of the crossover scale $T_*$, describing, e.g., the vanishing of the Kondo resonance half-width, is well fit by  
\be
\ln T_*\propto a_0-\frac{a_1}{\sqrt{B_{0,c}-B_{0}}}
\label{s1crossover}
\ee
as expected from the result Eq.\ (\ref{bfkms1scale}) for the Ohmic Bose-Fermi \textit{Kondo} model.

\begin{figure}
\begin{center}
\epsfig{file=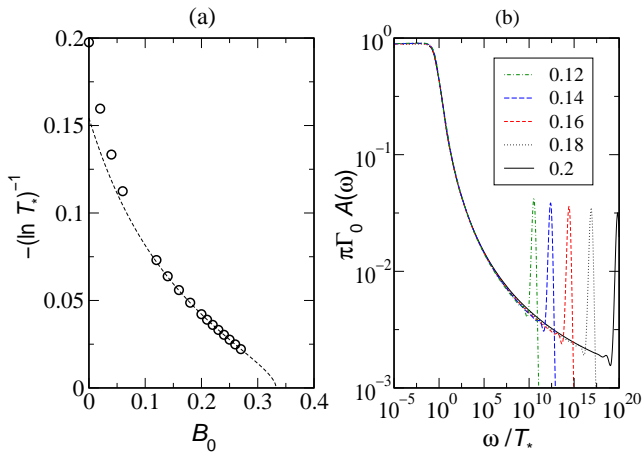, width=6.65cm, angle=270}
\end{center}
\caption{(Color online) (a) Vanishing of crossover scale $T_*$ for $B_0\ra B_{0,c}^-$, according to Eq.\ (\protect\ref{s1crossover}) (dashed line).  (b) Scaling of the single-particle spectrum $\pi\Gamma_0A(\w)$ for the dissipation strengths $B_0$ shown in the inset.  Though the values of $T_*$ in this data set vary across ten orders of magnitude, a single universal scaling curve [Eq.\ (\protect\ref{scalforms1}] is followed up to the nonuniversal scale set by the Hubbard satellite bands.}
\label{s1scaling}
\end{figure}

Third, the Kondo-phase spectra obey a simple scaling form
\be 
\pi\Gamma_0A(\w)=\phi_{s=1,K}\left(\frac{\w}{T_*}\right),   
\label{scalforms1}
\ee
as shown in Fig.\ \ref{s1scaling}(b).  The spectra there plotted, which have vastly different crossover scales (see, e.g., Fig.\ \ref{s1spec}), collapse to the single universal curve Eq.\ (\ref{scalforms1}), which form is followed up to the nonuniversal energy scale set by the Hubbard satellite bands, $\w_{\mbox{\ssz{H}}}/T_*$.  For the data shown, excellent data-collapse is seen up to $\w/T_*\simeq 10^{7}$.

We note that calculations of the $T$-matrix in Ref.\ \onlinecite{Borda:05}, discussed in the context of the quantum box, are in qualtitative agreement with the above.

\seceq
\section{Summary}
We have described an extension of Wilson's numerical renormalization group method appropriate for Bose-Fermi quantum impurity models.  We have focused in particular on the  Ising anisotropic Bose-Fermi Kondo model. The approach gives an excellent description of the critical properties of this model, which is of relevance to certain dissipative mesoscopic qubit devices and to a treatment of the anisotropic Kondo lattice model within an extension of dynamical mean-field theory.\cite{Glossop:06}
For sub-Ohmic bosonic bath exponents $0<s<1$ we find a continuous quantum phase transition between Kondo-screened and localized phases. The transition is governed by an interacting quantum-critical point, leading to hyperscaling of critical exponents and $\w/T$-scaling in the dynamics.   We have shown directly that the Kondo resonance in the spectral function is destroyed at the critical point. The case of Ohmic dissipation, which leads to a Kosterlitz-Thouless like transition, was also discussed.  

 While we have specifically studied the case of Ising symmetry bosonic couplings, the approach is not per se limited to that case.  It is straightfoward, albeit computationally more demanding, to add additional bosonic bath(s) as appropriate for the $xy$-anisotropic or isotropic  Bose-Fermi Kondo models. These cases will be discussed elsewhere.  Furthermore, our approach could be integrated  with the recently developed time-dependent NRG method\cite{Anders:05,Anders:06} to enable the study of  nonequilibrium dynamics for a wider range of quantum impurity systems.   Other models that might be tackled using the present approach include an Anderson-Holstein-type model (where the impurity charge density couples to a bosonic bath) relevant to the study of electron-phonon interactions in strongly correlated electron systems, and novel Bose-Fermi impurity problems that may arise in the treatment of ultracold  mixtures of bosonic and fermionic atoms trapped in optical lattices.    These applications are left for future study.

We thank Q.\ Si and R.\ Bulla for stimulating discussions regarding the present work.  Much of the computational work was performed using the University of Florida High Performance Computing Cluster; we thank C. Taylor in particular for technical support.  This work was supported in part by NSF Grant DMR-0312939.

\end{document}